\documentclass[10pt,journal,compsoc]{IEEEtran}
      
\usepackage[utf8]{inputenc}  
\usepackage[T1]{fontenc}                          
\usepackage{nicefrac}        
\usepackage{amsfonts}     
\usepackage{mathtools,lipsum,nccmath}

\usepackage{cuted}
\usepackage{xcolor}
\usepackage{graphicx}%
\usepackage{dcolumn} 
\usepackage{bm} 
\usepackage{braket} 
\usepackage{amsthm}

\usepackage[colorlinks=true, linkcolor=blue, citecolor=blue, urlcolor=blue]{hyperref}
   
\usepackage{url}   
\usepackage{multirow}

 \usepackage{environ} 
 \usepackage{caption}
 \usepackage{xifthen} 
\usepackage{hyperref}
\usepackage{booktabs}
 
\usepackage[ruled]{algorithm2e}
\usepackage{ragged2e}

\newtheorem{lemma}{Lemma}

\newtheorem*{theorem-non}{Theorem}
\newtheorem*{lemma-non}{Lemma}
\newtheorem*{corollary-non}{Corollary}
\newtheorem*{proposition-non}{Proposition}

\newcommand{\bx}{\bm{x}}
\newcommand{\by}{\bm{y}}
\newcommand{\bxi}{\bm{x}^{(i)}}

\newcommand{\rhoi}{\rho^{(i)}}
\newcommand{\byi}{\bm{y}^{(i)}}

\newcommand{\sigmai}{\sigma^{(i)}}
\newcommand{\bW}{\bm{w}}

\newcommand{\PP}{\mathbb{P}}

\newcommand{\tilderhoi}{\widetilde{\rho}^{(i)}}

\newcommand{\bzi}{\bm{z}^{(i)}}

\DeclareMathOperator{\CNOT}{CNOT}

\DeclareMathOperator{\hatrho}{\hat{\rho}}
\DeclareMathOperator{\hatrhoi}{\hat{\rho}^{(i)}}

\DeclareMathOperator{\GHZ}{\text{GHZ}}

\DeclareMathOperator{\CI}{\text{CI}}
\DeclareMathOperator{\Tr}{Tr}
\DeclareMathOperator{\Loss}{\mathcal{L}}
\DeclareMathOperator{\loss}{\mathtt{L}}

\usepackage{pifont}

\title{ShadowNet for Data-Centric Quantum System Learning}

\author{Yuxuan~Du, Yibo~Yang, Tongliang Liu,  Zhouchen Lin,~\IEEEmembership{Fellow,~IEEE}, David A. Clifton, \\Bernard Ghanem,  and~Dacheng~Tao,~\IEEEmembership{Fellow,~IEEE}
\IEEEcompsocitemizethanks{\IEEEcompsocthanksitem Y. Du is with the College of Computing and Data Science, and the School of Physical and Mathematical Sciences, Nanyang Technological University, 639798, Singapore. 
\IEEEcompsocthanksitem Y. Yang is with School of Artificial Intelligence, Shanghai Jiao Tong University and King Abdullah University of Science and Technology,  Thuwal, 23955 – 6900, Kingdom of Saudi Arabia.   
\IEEEcompsocthanksitem T.~Liu is with the School of Computer Science, the Faculty of Engineering, in the University of Sydney, Darlington, NSW 2008, Australia.
\IEEEcompsocthanksitem Z.~Lin is with the State Key Lab of General AI, School of Intelligence Science and Technology, Peking University.
\IEEEcompsocthanksitem D. A. Clifton is with the Institute of Biomedical Engineering, Department of Engineering Science, the University of Oxford, Oxford, OX1 4BH, UK.
\IEEEcompsocthanksitem B.~Ghanem is with  King Abdullah University of Science and Technology,  Thuwal, 23955 – 6900, Kingdom of Saudi Arabia.   
\IEEEcompsocthanksitem D. Tao is with the College of Computing and Data Science, Nanyang Technological University, 639798, Singapore.  
} 
\thanks{This project is supported by the National Research Foundation, Singapore, under its NRF Professorship Award No. NRF-P2024-001. Y. Du is supported by Agency for Science, Technology and Research (A*STAR), Singapore, under award number H25-MRO3488. B. Ghanem is supported by funding from King Abdullah University of Science and Technology (KAUST) - Center of Excellence for Generative AI, under award number 5940. (Corresponding author: Dacheng Tao.)}
}

\begin{document}

\IEEEtitleabstractindextext{
\begin{abstract}
\justifying
Understanding the dynamics of large quantum systems is hindered by the curse of dimensionality. Statistical learning offers new possibilities in this regime through neural network protocols and classical shadows, while both methods have limitations: the former suffers from incompatible dataset construction rules, resulting in substantial computational demands for data collection when addressing different tasks; the latter lacks the ability to distill knowledge from prior data to enhance subsequent learning endeavors. In this study, we propose a data-centric learning paradigm combining the strengths of these two approaches to advance quantum system learning (QSL).  Central to our paradigm lies a unified dataset construction rule, achieved by classical shadows along with other easily obtainable information of quantum systems. To illustrate our approach, we present ShadowNet, implemented under both convolutional and attention mechanisms, to efficiently and faithfully tackle two pivotal QSL tasks: quantum state tomography (QST) and direct fidelity estimation (DFE). Numerical simulations on QST and DFE up to 60 qubits validate the efficacy of our proposal, showcasing how ShadowNet advances classical shadows with limited state copies, and highlighting how the varied neural networks impact the performance. Our work underscores the immense potential of a data-centric approach in comprehending novel and large quantum systems.
\end{abstract}

}

\maketitle
 
\IEEEdisplaynontitleabstractindextext
 
\IEEEpeerreviewmaketitle

\maketitle

\IEEEraisesectionheading{\section{Introduction}\label{sec:introduction}}
\IEEEPARstart{T}{he} precise characterization of quantum systems holds paramount significance in the development, validation, and evaluation of emerging quantum technologies \cite{eisert2020quantum,gebhart2023learning}. However, the rapid progress of quantum technologies presents an escalating challenge in fully describing modern quantum devices, since a general $N$-qubit state tomography necessitates exponential state replicas with $N$ for a reliable estimation \cite{gross2010quantum,blume2010optimal,flammia2012quantum,sugiyama2013precision,haah2016sample,o2016efficient,kueng2017low}. Fortunately, many physical scenarios offer a ray of hope as the structural information about the target systems is known, allowing scenario-specific models to characterize and represent large-scale quantum systems efficiently. As a result, great efforts have been made to designing effective models for quantum system characterization, encompassing tasks such as low-entangled state reconstruction \cite{cramer2010efficient,landon2012practical,lanyon2017efficient}, randomized benchmarking \cite{knill2008randomized,magesan2012efficient,helsen2022general}, direct fidelity estimation \cite{flammia2011direct,huang2020predicting,elben2022randomized}, Hamiltonian learning \cite{granade2012robust,bairey2019learning,huang2023learning}, and self-testing \cite{mayers2003self,montanaro2013survey,vsupic2020self}. Despite their diversity across specific tasks, all models share a common target: \textit{fast} acquire the desired information of the quantum system and \textit{minimize} the number of required state copies. 

A leading solution in large-scale quantum system characterization and representation is shadow tomography, standing on the fact that many practical QSL tasks only require accurate predictions of specific properties of quantum systems, rather than their full classical descriptions. Myriad algorithms have been proposed to complete shadow tomography \cite{aaronson2018online,chen2022adaptive,lumbreras2022multi} in recent years. Among them, the most viable one is classical shadows \cite{huang2020predicting}, requiring only polynomial state copies to predict an exponential number of target functions. Statistical learning algorithms are another leading candidate to tackle quantum system characterization and representation tasks, a.k.a, quantum system learning (QSL), due to their intrinsic data-driven nature  \cite{aaronson2007learnability,mohri2018foundations}. A prominent approach in this context are shadow tomography~\cite{aaronson2018shadow} and neural network based QSL (NN-QSL) \cite{torlai2018neural, gao2018experimental,zhu2022flexible,wang2022predicting}. In particular, NN-QSL makes use of the strong power of deep neural networks (DNNs) to learn a class of quantum states with similar structures by extracting their hidden features.  Over the past years, varied architectures of DNNs have been proposed to tackle different QSL tasks, capitalizing on the enhanced efficiency \cite{carrasquilla2019reconstructing,torlai2019integrating,xin2019local,ahmed2021quantum,quek2021adaptive,morawetz2021u,cha2021attention,lange2022adaptive,koutny2022neural,schmale2022efficient,zhang2022efficient,an2023unified} and generalization ability \cite{wu2023quantum,zhang2021direct,smith2021efficient,zhong2022quantum,lohani2023dimension}.

Despite their significant potential, both classical shadows and NN-QSL exhibit notable limitations when aiming to comprehensively characterize a quantum system. Recall that each QSL approach can only reveal partial properties of a quantum system, and obtaining a full characterization typically necessitates a combination of multiple QSL tasks \cite{gebhart2023learning}. Nevertheless, classical shadows and their variants might even yield inferior performance when employed for a single QSL task. This is because classical shadows and their variants are fundamentally designed to estimate specific properties of individual quantum states \cite{elben2022randomized}, without the capacity of harnessing knowledge acquired from a class of states to reduce the number of state copies required for future estimations. In practical scenarios where estimating a specific property for a large volume of quantum states is required, and the number of measurements is limited, the average number of measurements allocated to each state becomes exceedingly constrained,  leading to a significant increase in estimation error of classical shadows and its variants.

On the other hand, the prior literature on NN-QSL has predominantly adhered to a \textit{model-centric} approach. Within this framework, the primary focus has been on crafting task-specific learning models, involving decisions about neural network architectures and objective functions. Unfortunately, this focus often occurs with limited consideration for the interconnections between different QSL tasks. As a result, the dataset construction rules are usually tailored to each individual QSL task, resulting in inherent incompatibilities between them. However, due to the resource-intensive nature of data collection from quantum devices, these incompatibilities are compounded by the restricted amount of training data available for each task, ultimately leading to diminished learning performance. 

The limitation of classical shadows and model-centric QSL interpreted above motivates us to ask a fundamental question: \textit{Is there a solution capable of efficiently and effectively characterizing a quantum system across a spectrum of QSL tasks?}
 
In this study, we respond affirmatively to the question posed above by introducing the concept of \textit{data-centric QSL}. Different from model-centric QSL that concentrates on designing task-specific learning models, data-centric QSL emphasizes the pivotal role of the dataset in enhancing both the efficiency and performance of versatile NN-QSL protocols to address multiple QSL tasks. 

A major challenge in data-centric QSL is designing a unified dataset construction rule capable of accommodating diverse QSL tasks. This challenge is fundamentally rooted in the curse of dimensionality characterizing quantum systems, which renders the storage of complete quantum state descriptions an infeasible endeavor, especially as the qubit count scales. While a potential avenue entails preserving partial information about quantum states to construct their classical descriptions, the selection of which information to retain for effective learning across various QSL tasks remains obscure. To overcome this challenge, we propose the first dataset construction rule toward data-centric QSL, achieved by integrating classical shadows with readily available information pertaining to the quantum system under investigation. Attributed to the memory efficiency of classical shadows and theoretical evidence of its capability to manipulate myriad QSL tasks, data-centric QSL executed on our proposed dataset can confer advantages upon diverse tasks, particularly those that classical shadows can effectively address.

To elucidate the concept of data-centric QSL, we further design a general learning framework, dubbed \textit{ShadowNet}, which leverages the proposed dataset and NN-QSL models to characterize a quantum system across multiple QSL tasks. The fusion of classical shadows and NN-QSL protocols effectively mitigates their respective limitations and warrants the efficiency and generalization of ShadowNet. Moreover, ShadowNet embraces two extra attractive properties. Firstly, ShadowNet can seamlessly adapt to a variety of neural network architectures, making it a highly flexible tool that can collaborate effectively with existing NN-QSL protocols to enhance its capabilities. Secondly, the inclusion of classical shadows as data features allows ShadowNet to quantify predictive confidence for unseen examples, a crucial property that underpins ShadowNet's reliability. We have also conducted an analysis of how predictive confidence varies with the number of classical shadow measurements.  

For concreteness, we illustrate how to apply ShadowNet to solve two pivotal QSL tasks: quantum state tomography (QST) and direct fidelity estimation (DFE) under both convolutional and attention mechanisms. Considering that the format of classical shadows is not compatible with DNNs, a data feature engineering process is required to transform the raw data into a format amenable to DNN processing. To address this, we design several strategies for data feature engineering. Notably, the proposed formalism and learning strategy can be extended to tackle other substantial QSL tasks, which may be of independent interest. 
 
The main contributions of this work are summarized as follows.
\begin{itemize}
	\item We lay the foundation for data-centric quantum system learning rooted in the principles of classical shadows. Our proposal offers a promising avenue for generating a unified database to advance a wide array of quantum system learning tasks, particularly those that classical shadows can effectively address. 
	\item We present a novel learning framework, dubbed ShadowNet,  to \textit{efficiently},  \textit{faithfully}, and  \textit{generalizably} solve a wide range of QSL tasks by utilizing the unified dataset. We perform theoretical analysis to show how the predictive faithfulness of ShadowNet depends on the shot number of classical shadows.   
\item We conduct systematic simulations to exhibit the efficacy of ShadowNet when tackling QST and DFE tasks under both convolutional and attention mechanisms. In the context of reconstructing ground states of spin systems, ShadowNet exhibits an order of magnitude enhancement over classical shadows utilizing the same measurement count. In the realm of direct fidelity estimation for 60-qubit noisy GHZ states, ShadowNet achieves near-perfect estimation using just 200 training examples.  
\end{itemize}

\subsection{Related works}

 Prior literature related to our work can be classified into three categories, which are classical shadows, neural-network-based quantum state tomography (NN-QST), and other neural-network-based quantum system learning (NN-QSL) protocols. In what follows, we elucidate the connections and distinctions between our work and previous studies falling within each of these categories.

\noindent\textbf{Classical shadows}. The primary objective of classical shadows is to efficiently estimate the expectation values of relevant observables using only a limited number of measurements \cite{huang2020predicting}. Since the inception of classical shadows, researchers have been actively investigating advanced measurement protocols \cite{nguyen2022optimizing,zhou2023performance,ippoliti2023classical} and de-randomization methods \cite{huang2021efficient,nakaji2023measurement} that can be efficiently implemented on modern quantum platforms while reducing the number of required measurements for accurate estimation. Additionally, extensive research has been conducted on the various applications of classical shadows, such as solving quantum many-body problems \cite{huang2022quantum,huang2022provably}, quantum error mitigation \cite{hu2022logical,jnane2023quantum,seif2023shadow}, and improving variational quantum algorithms \cite{sack2022avoiding,boyd2022training,du2022demystify}.

Our work complements the existing literature in this domain. Specifically, methods that enhance the capabilities of classical shadows can benefit ShadowNet. Furthermore, ShadowNet can be extended to tackle QSL tasks that are amenable to classical shadows' approaches. This complementary role arises from our focus on exploring the memory efficiency and theoretical guarantees of classical shadows in forming the training dataset, which is subsequently learned by deep neural networks to distill the latent knowledge of the studied quantum system.

\noindent\textbf{NN-QST}. The seminal work on NN-QST was proposed by Torlai et al \cite{torlai2018neural}. Since then, this topic has evolved into two distinct research lines  \cite{smith2021efficient}. The first line involves explicit state reconstruction, where the output of the NN represents the density matrix of the target quantum state \cite{ahmed2021quantum,cha2021attention,ma2023attention}. The second line focuses on implicit QST, wherein NN is optimized to emulate the behavior of a given quantum state. This research line comprises two sub-paradigms. In the first sub-paradigms, the NN takes a specified measurement basis as input and produces the corresponding measurement probabilities as output  \cite{schmale2022efficient,smith2021efficient}. The second sub-paradigm treats the neural network as a sampler, generating outputs in the form of computational basis samples drawn from the probability distribution corresponding to the true quantum state \cite{torlai2018neural,an2023unified}.

A fundamental distinction between our work and prior NN-QST studies lies in our emphasis on the significance of training examples (data-centric AI), rather than solely focusing on diverse DNN architectures for addressing QST (model-centric AI). Additionally, unlike previous approaches that treat \textit{explicit} QST as an optimization problem, where the optimized DNN can only reconstruct a single quantum state, our work takes a concrete stride towards solving explicit QST in a more generalizable manner.

\noindent\textbf{NN-QSL}. Besides QST, NN-based protocols have been applied to tackle other QSL problems, including fidelity estimation \cite{zhang2021direct}, energy estimation \cite{zhu2022flexible}, entropy estimation \cite{nir2020machine}, cross-platform verification \cite{xiao2022intelligent,qian2024multimodal}, and similarity testing \cite{wu2023quantum}. As with NN-QST, a common feature of prior studies is model-centric, as the main focus is to design novel neural architectures and objective functions. By contrast, our work places a greater emphasis on presenting a unified construction rule of datasets to facilitate a broad class of QSL tasks.
 
\smallskip
The outline of the rest of the paper is as follows. In Sec.~\ref{append:sec:background}, for self-consistency, we introduce elementary concepts of quantum computing, QST, DFE, classical shadows, and neural networks designed for quantum system learning. Then, in Sec.~\ref{sec:shadownet-all}, we present the dataset construction rule for data-centric QSL and a general framework of ShadowNet and its detailed implementations when applied to QST and DFE. Subsequently, in Sec.~\ref{sec:num-sim}, we conduct systematic experiments to comprehend the performance of ShadowNet in the tasks of QST and DFE. Last, we discuss potential avenues for future research in Sec.~\ref{sec:discussion}.
  
\section{Preliminary}\label{append:sec:background}
In this section, we provide the necessary background of classical shadows and deep learning models to better understand ShadowNet. With this aim, we first introduce the basic concepts of quantum computing, quantum state tomography, and direct fidelity estimation. Next, we recap classical shadows and explain how to use this technique to estimate linear functions. We last review the attention mechanism and Cholesky decomposition layer, as the core ingredients to construct ShadowNet.   

\subsection{Quantum computing}
The fundamental unit in quantum computing is a qubit, which refers to a two-dimensional vector. Under Dirac notation, a qubit state is denoted by  $\ket{\bm{\alpha}}=a_0\ket{0}+a_1\ket{1}\in\mathbb{C}^2$, where $\ket{0}=[1,0]^{\top}$ and $\ket{1}=[0,1]^{\top}$ specify two unit bases, and the coefficients $a_0, a_1 \in \mathbb{C}$ satisfy $|a_0|^2+|a_1|^2=1$. An $N$-qubit state is denoted by $ \ket{\Psi}=\sum_{i=1}^{2^N}a_i\ket{i} \in \mathbb{C}^{2^N}$, where $\ket{i}\in\mathbb{R}^{2^N}$ is the unit vector whose $i$-th entry being 1 and other entries are 0, and $\sum_{i=0}^{2^N-1}|a_i|^2=1$ with $a_i\in\mathbb{C}$. We sometimes denote $\ket{\bm{a}}\ket{\bm{b}}$ as $\ket{\bm{a}}_A\ket{\bm{b}}_B$, which means that the qubits $\ket{\bm{a}}_A$ ($\ket{\bm{b}}_B$) are assigned in the quantum register $A$ ($B$). Besides Dirac notation, the density matrix can be used to describe more general qubit states. For example, the density matrix of the state $\ket{\Psi}$ is $\rho=\ket{\Psi}\bra{\Psi}\in \mathbb{C}^{2^N\times 2^N}$. For a set of qubit states $\{p_i, \ket{\psi_i} \}_{i=1}^m$ with $p_i>0$, $\sum_{i=1}^{m}p_i=1$, and $\ket{\psi_i}\in \mathbb{C}^{2^N}$ for $\forall i\in [m]$, its density matrix is $\rho=\sum_{i=1}^{m} p_i\rho_i$ with $\rho_i=\ket{\psi_i}\bra{\psi_i}$ and $\Tr({\rho})=1$. A quantum state $\rho$ is separable if it can be written as a convex combination of product states $\rho= \sum_j p_j \rho^{(j)}_1\otimes\cdots\otimes\rho^{(j)}_N$ with $p_j\ge 0, \sum_j p_j=1.$ Otherwise, $\rho$ is entangled.

Quantum gates refer to unitary transformations and serve as the computational toolkit for quantum circuit models, i.e., an $N$-qubit gate $U\in\mathcal{U}(2^N)$ obeys $UU^{\dagger}=U^{\dagger}U=\mathbb{I}_{2^N}$, where $\mathcal{U}(\cdot)$ stands for the unitary group. Typical single-qubit quantum gates include the Pauli gates, which can be written as Pauli matrices, i.e., $X=\begin{psmallmatrix} 
	0 & 1 \\
				1 & 0 \\
\end{psmallmatrix}
$, $Y= \begin{psmallmatrix} 
	0 & -i \\
				i & 0 \\
\end{psmallmatrix}$, and $Z= \begin{psmallmatrix} 
	1 & 0 \\
				0 & -1 \\
\end{psmallmatrix}$. 
The more general quantum gates are their corresponding rotation gates $R_X(\theta)=e^{-i\frac{\theta}{2}X}, R_Y(\theta)=e^{-i\frac{\theta}{2}Y}$, and $R_Z(\theta)=e^{-i\frac{\theta}{2}Z}$ with a tunable parameter $\theta$.  Moreover, a multi-qubit gate can be either an individual gate (e.g., CNOT gate with $\CNOT=\begin{psmallmatrix} 1 & 0 & 0 & 0 \\ 0 & 1 & 0 & 0 \\ 0 & 0 & 0 & 1 \\  0 & 0 & 1 & 0 \end{psmallmatrix}$) or a tensor product of multiple single-qubit gates.
 
Quantum measurements are employed to extract quantum information of the evolved quantum state into the classical form. An observable is represented by a Hermitian operator $H$. If $H$ has spectral decomposition $H=\sum_\lambda \lambda \Pi_\lambda$ with $ \Pi_\lambda$ being the spectral projector, then measuring $H$ on a state $\rho$ returns outcome $\lambda$ with probability $\mathrm{Pr}(\lambda)=\mathrm{Tr}(\Pi_\lambda\rho)$. The expectation value of this measurement is $\mathbb{E}[H]=\mathrm{Tr}(\rho H)$.

\subsection{Quantum state tomography}\label{sec:pre-QST}
Quantum state tomography (QST) is the process by which the density matrix of a quantum state is reconstructed using measurements on a set of $M$ identical state copies. Formally, for an $N$-qubit system, let $\{\Pi_i\}$ be a tomographically complete measurement such that $\forall i$, $\Pi_i$ is a positive semidefinite matrix  and $\sum_i\Pi_i=\mathbb{I}$. Applying the measurement $\{\Pi_i\}$ to the state $\rho$, the probability of observing outcome $i$ depends on $\rho$ and is described by Born’s rule, i.e., $\Pr(i|\rho) = \Tr(\rho \Pi_i)$. The probabilities $\{\Pr(i|\rho)\}$ can be estimated by frequencies with $f(i,M)=M_i/M$, where $M_i$ refers to the number of times outcome $i$ was observed. Supported by the central limit theorem, these frequencies converge to the true probabilities when $M \rightarrow\infty$.  The task of QST considers using the collected $\{f(i,M)\}$ to recover a state $\widetilde{\rho}$ that estimates $\rho$ with a low error bound. Conventional QST methods typically optimize each state independently, disregarding the potential relationships among states that may exhibit similar structures and sample from the same underlying distribution $\mathbb{D}$.  

\subsection{Direct fidelity estimation}

Direct fidelity estimation (DFE) is a crucial measure in assessing the quality of quantum states prepared by quantum hardware  \cite{flammia2011direct}. Mathematically, DFE measures the similarity between the desired pure state $\rho$ and the actual state $\sigma$, i.e., 
\begin{equation}
F_Q=\Tr(\sigma\rho).	
\end{equation}
In the quest to certify large-scale quantum devices, the primary objective of DFE algorithms is to obtain accurate estimations of $F_Q$ with few measurements $M$ and efficient post-processing \cite{kliesch2021theory}. Similarly to QST, conventional fidelity estimators focus on individual states, neglecting the inherent relationships among a set of states and thereby facing challenges in terms of either sample complexity or post-processing runtime cost. 

It is noteworthy that DFE is not the only fidelity metric relevant for quantum system certification. Other important measures include Uhlmann fidelity~\cite{nielsen2010quantum} that quantifies the closeness between arbitrary quantum states, and cross-platform fidelity~\cite{elben2020cross} that evaluates the similarity of output states produced by different quantum processors when executing the same program.

\subsection{Classical shadows}\label{sec:pre:shadow}
 Classical shadows represent a computationally and memory-efficient approach for storing quantum states on classical computers, primarily used for estimating the expectation values of local observables \cite{huang2020predicting}. The fundamental principle of classical shadows lies in the `measure first and ask questions later' strategy. In this subsection, we outline the utilization of classical shadows to estimate linear functions under Pauli-based measurements. Interested readers can refer to tutorials and surveys \cite{elben2022randomized,huang2022learning} for more comprehensive details.

The general scheme of classical shadows for an unknown state $\rho$ under Pauli-based measurements is to repeat the following procedure $M$ times. At the $m$-th time, the state $\rho$ is first operated with a unitary $U_m$ randomly sampled from the predefined unitary ensemble $\mathcal{U}$ and then each qubit is measured in the Z basis to obtain an $N$-bit string denoted by $\bm{b}_m$. Define the $m$-th snapshot as $U_m^{\dagger}\ket{\bm{b}_m}\bra{\bm{b}_m}U_m$. There exists a linear map $\mathcal{M}(\cdot)$ satisfying 
\begin{equation}
	\mathcal{M}(\rho) =\mathbb{E}_{U\sim \mathcal{U}}\mathbb{E}_{\bm{b}\sim \PP(\bm{b})} U^{\dagger}\ket{\bm{b}}\bra{\bm{b}} U,   
\end{equation} 
where $\PP(b)=\bra{b}U\rho U^{\dagger}\ket{b}$. Such a linear map implies that the unknown state $\rho$ can be formulated as 
\begin{equation}
	\rho =  \sum_b \int dU \mathcal{M}^{-1}\Big(U^{\dagger}\ket{b}\bra{b} U\Big) \bra{b}U\rho U^{\dagger}\ket{b}.
\end{equation} 
In other words, the state $\rho$ can be estimated by sampling the snapshot with $M$ times following the  $\PP(b)$, i.e., the estimated state of $\rho$  is 
 \begin{equation}
	\hat{\rho} = \frac{1}{M}\sum_{m=1}^M \hatrho_m,~\text{with}~\hatrho_m=\mathcal{M}^{-1}(U_m^{\dagger}\ket{\bm{b}_m}\bra{\bm{b}_i} U_m).
\end{equation}

As pointed out in Ref.~\cite{huang2020predicting}, the random Pauli basis is not only experimentally friendly, but also enables a succinct form of the classical shadow. When Pauli-based measurements are adopted, it is equivalent to setting the unitary ensemble $\mathcal{U}$ as single-qubit Clifford gates, i.e., $U_m=U_{1,m}\otimes  \cdots U_{j,m} \cdots \otimes U_{N,m}\sim \mathcal{U}=\CI(2)^{\otimes N}$ with uniform weights. In this case, the inverse snapshot has a succinct form, i.e.,  
\begin{eqnarray}\label{append:SM-A-CS-form}
	\hatrho_m && = 
	\bigotimes_{j=1}^N  \left(3U_{j,m}^{\dagger} |\bm{b}_{j,m}\rangle\langle \bm{b}_{j,m}|U_{j,m} - \mathbb{I}_2 \right). 
\end{eqnarray} 

The tensor product form of classical shadows in Eq.~(\ref{append:SM-A-CS-form}) allows an efficient procedure to predict linear functions. A typical instance is estimating the expectation value $\Tr(\rho O)$ with $O$ being the local observable. Mathematically, suppose the local observable is a Pauli string, i.e., $O=P_1\otimes ...P_i...\otimes P_N$ with $P_i\in\{X,Y,Z,\mathbb{I}\}$ for $\forall i \in [N]$, the estimation of  classical shadows is
\begin{equation}
	\Tr\left(\hatrho O  \right) = \frac{1}{M}\sum_{m=1}^M\prod_{j=1}^N \Tr\left(\big(3U_{j,m}^{\dagger} |\bm{b}_{j,m}\rangle\langle \bm{b}_{j,m}|U_{j,m} - \mathbb{I}_2 \big)P_{j}\right), 
\end{equation}  
 which is memory and computation efficient.

\subsection{Attention mechanism}\label{sec:layers-DNN-QSL}
The self-attention mechanism \cite{vaswani2017attention} serves as a fundamental block of the Transformer, an advanced deep neural network architecture that has achieved remarkable performance across various applications such as image processing \cite{dosovitskiy2020image} and natural language processing \cite{Devlin2019BERT}. The core idea behind the self-attention mechanism is to establish relationships between different input segments through the use of dot product computations and dynamically pay attention to only certain parts of the input that help in performing the task at hand effectively. In this subsection, we provide a brief introduction to its implementation. 

Denote the segmentation of the input example $\bx$ as $\{\bx_s\}_{s=1}^S$, e.g., an image with $8\times 8$ pixels can be segmented into $S=4$ pieces with $\bx_s$ being the $16$-dimensional vector. Self-attention first calculates a normalized dot product between all pairs of input vectors in $\{\bx_s\}_{s=1}^S$. Mathematically, the \textit{query}, \textit{key}, and \textit{value} of $\bx_s\in \mathbb{R}^{p}$ for $\forall s \in [S]$ are defined as
\begin{equation}\label{eqn:append-A-mechanism-attention}
	\bm{q}_s=W_q\bx_s \in \mathbb{R}^{d_k}, ~ \bm{k}_s=W_k\bx_s \in \mathbb{R}^{d_k}, ~
	\bm{v}_s=W_v\bx_s \in \mathbb{R}^{d_v},
\end{equation}
where $W_q\in \mathbb{R}^{d_k\times p}$, $W_k\in \mathbb{R}^{d_k\times p}$, and $W_v\in \mathbb{R}^{d_v\times p}$ are trainable. The normalized correlation between $\bx_s$ and each segmentation $\bx_{s'}$ with $s'\in [S]$ yields
\begin{equation}
	\omega_{ss'} = \text{softmax}_{s'}\!\left(\bm{q}_s^{\top} \bm{k}_{s'}/\sqrt{d_k}\right)
	= \frac{\exp(\bm{q}_s^{\top} \bm{k}_{s'} /\sqrt{d_k} )}
	       {\sum_{j=1}^{S}\exp(\bm{q}_s^{\top} \bm{k}_{j} /\sqrt{d_k})},
\end{equation}
which satisfies $\omega_{ss'}\ge 0$ and $\sum_{s'=1}^{S}\omega_{ss'}=1$. Then, a new representation of $\bx_s$ is formed by aggregating the values $\{\bm{v}_{s'}\}_{s'=1}^{S}$ with the weights $\{\omega_{ss'}\}_{s'=1}^{S}$, i.e.,
\begin{equation}
	\bm{z}_s = \sum_{s'=1}^S \omega_{ss'} \bm{v}_{s'} \in \mathbb{R}^{d_v}.
\end{equation}
In this way, the resulting representation $\bm{z}_s$ is dominated by the value $\bm{v}_{s'}$ of the segmentation $\bx_{s'}$ that receives the largest attention weight $\omega_{ss'}$.

\subsection{Cholesky decomposition layer}\label{sec:pre-cholesky} 
The Cholesky decomposition \cite{banaszek1999maximum} has also been exploited in the prior literature of NN-QST \cite{ahmed2021quantum,koutny2022neural}. This operation layer does not require any optimization; instead, it applies Cholesky factorization to transform the input matrix into a physical state, i.e., a positive semidefinite matrix with the identity trace norm. Mathematically, this layer takes the unconstrained output of the fully-connected layer and forges it into a lower triangular complex-valued matrix $T$ with real entries on the diagonal. Then, its output yields $TT^{\dagger}/\Tr(TT^{\dagger})$, which is a physical state.

\section{Implementation of ShadowNet}\label{sec:shadownet-all}
In this section, we begin by introducing a unified dataset construction rule for data-centric quantum system learning (QSL). Following this, we present a general learning framework, dubbed ShadowNet, for utilizing the proposed dataset to advance multiple QSL tasks simultaneously. To provide a concrete understanding, we subsequently exhibit the implementation specifics of ShadowNet when applied to accomplish two significant QSL tasks: quantum state tomography (QST) and direct fidelity estimation (DFE), employing two distinct neural network architectures.

\subsection{Dataset construction rule for data-centric QSL}\label{sec:gene-frame-shadownet}
\begin{figure}
	\centering
\includegraphics[width=0.48\textwidth]{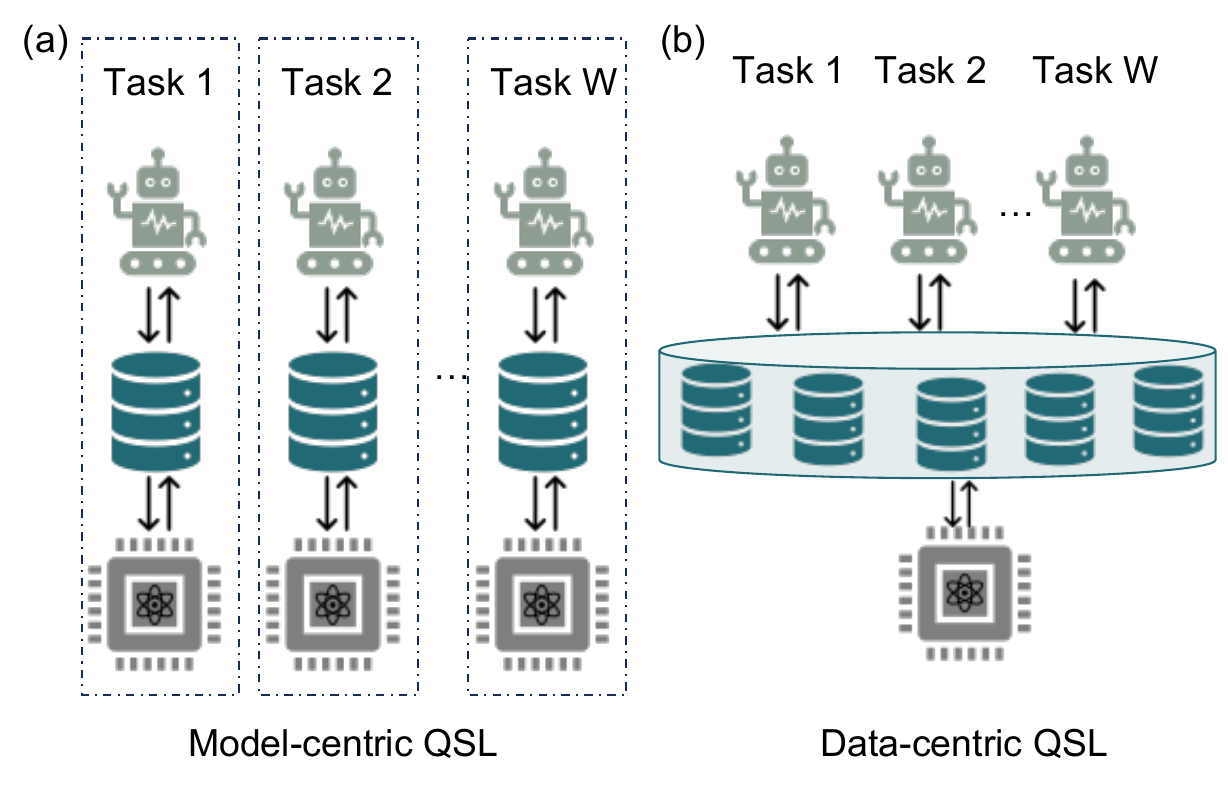}
\caption{\footnotesize{\textbf{Model-centric QSL versus data-centric QSL}.  The model-centric QSL centers on exploring the difference among diverse deep learning models for a specific QSL task.  In contrast, the data-centric QSL approach revolves around the creation of a unified dataset that empowers learning models to tackle a broad spectrum of QSL tasks by training on this versatile dataset.} }
\label{fig:data-vs-model}
\end{figure}

Data-centric QSL underscores the paramount role of enhancing datasets to bolster both the efficiency and performance of learning models when processing a broad spectrum of QSL tasks. As shown in Fig.~\ref{fig:data-vs-model}, this concept complements prior model-centric QSL, which centers on designing task-specific learning models. To be more concrete, the model-centric approach involves the unique design of dataset construction rules tailored to each specific task, rendering these datasets non-reusable for other tasks. Nevertheless, given the resource-intensive nature of collecting data from quantum machines, employing this approach for characterizing a quantum machine across \textit{multiple} QSL tasks may encounter inferior performance due to limited training data. In this regard, here we explore and advocate a data-centric QSL approach, addressing this restriction by creating a unified dataset to tackle multiple QSL tasks.

Despite the promising prospect, the construction of such datasets is challenging due to the curse of dimensionality inherent to quantum systems. On the one hand, utilizing the complete quantum state description to forge the dataset is infeasible as computational demands and classical memory requirements exponentially grow with the qubit count. On the other hand, preserving partial information of the quantum state to forge the dataset offers a potential solution. Yet, it remains unclear which specific information should be retained to ensure it is both memory- and computation-efficient while carrying sufficient information for effective learning. Taken together, there are three fundamental criteria for the data-centric QSL:
\begin{enumerate}
	\item memory efficient;
	\item  computation efficient;
	\item  convey sufficient information to solve multiple QSL problems.
\end{enumerate}

\begin{figure*}  
	\centering
\includegraphics[width=0.95\textwidth]{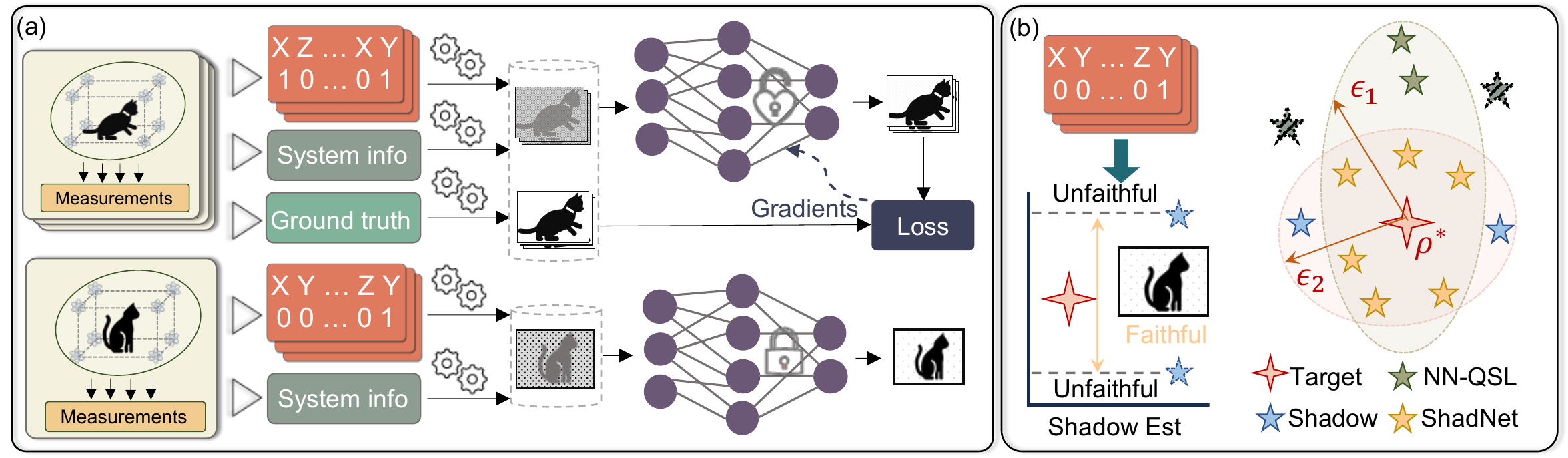}
\caption{\footnotesize{\textbf{The scheme of ShadowNet}.  (a) The basic mechanism of ShadowNet. The upper panel illustrates the training procedure. ShadowNet constructs the training dataset in which each example consists of classical shadows, system information, and ground truth. The classical shadows and system information should be preprocessed before being fed into a handcrafted deep neural network (DNN), highlighted by the dashed cylinder. The handcrafted DNN is optimized for $T$ epochs to minimize the predefined loss. The lower panel depicts the inference procedure. Given a new input, the same preprocessing rule is applied, and then the processed data is fed into the trained DNN.   (b) The left panel shows the faithfulness evaluation of ShadowNet. The predictive faithfulness of ShadowNet can be effectively examined by its shadow estimation. The right panel depicts the relation of ShadowNet, classical shadows, and NN-QSL.   The labels `Target', `NN-QSL', `Shadow', and `ShadNet' refer to the target result, the outputs of NN-QSL, classical shadows, and ShadowNet, respectively. Although NN-QSL enables the estimation error of training data below $\epsilon_1$, the predictive faith of new states is unwarranted, highlighted by the dashed green stars. The estimation error of classical shadows $\epsilon_2$  depends on the number of measurements on each state. ShadowNet incorporates the strengths of classical shadows and NN-QSL to faithfully predict unseen states using very few copies. }}
\label{fig:scheme}
\end{figure*}

To conquer the above issue, here we propose a unified and efficient dataset construction rule towards data-centric QSL. The central idea of our proposal is to integrate the classical shadows introduced in Sec.~\ref{sec:pre:shadow} with readily accessible information about the quantum system to form the dataset. Suppose that we intend to characterize an $N$-qubit system across $W$ distinct QSL tasks. The mathematical expression of the proposed dataset for these $W$ tasks is 
\begin{equation}
\mathcal{D}_{\text{QSL}}=\{(\mathcal{C}^{(i)}, \bzi), (\by^{(i,1)},...,\by^{(i,w)}, ..., \by^{(i,W)})\},
\end{equation}
where $\mathcal{C}^{(i)}= \{(U_{j,m}^{(i)}, b_{j,m}^{(i)})\}_{j,m=1}^{N,M}$ refers to the Pauli-based classical shadows of the $i$-th quantum state with $M$ being the shot number, and $U_{j,m}^{(i)}$ being the $m$-th Pauli operator and $b_{j,m}^{(i)}\in\{0, 1\}$ being the measured bit on the $j$-th qubit; the notation $\bzi$ refers to the readily accessible information about the quantum system, e.g., the qubit connectivity or the noise level of the quantum system; the notation $\by^{(i,w)}$ stands for the label of the $i$-th quantum state for the $w$-th task. Armed with this unified dataset $\mathcal{D}_{\text{QSL}}$, prior learning models designed for the explored $W$ QSL tasks can be effectively harnessed to acquire knowledge from this dataset and select the model demonstrating optimal performance.

Let us end this subsection by elaborating on how $\mathcal{D}_{\text{QSL}}$ meets the three fundamental criteria for data-centric QSL.  

\noindent\textbf{For Criterion 1)}. In the measure of the memory cost, both $\mathcal{C}^{(i)}$ and $\bzi$ can be tailored to polynomially scale with the qubit count $N$. Moreover, for most QSL tasks, the label $\by^{(i,w)}$ refers to a scalar, where representative examples include fidelity estimation, entanglement entropy quantification, state discrimination, and ground energy estimation. While one exception is explicit quantum state tomography such that the dimension of the label exponentially scales with $N$,  the magnitude of $N$ of practical interest is generally set to be small. In consequence, the constructed dataset $\mathcal{D}_{\text{QSL}}$ is a well-suited solution to satisfy Criterion 1). 

\noindent\textbf{For Criterion 2)}. The computation cost pertains to the cumulative number of measurements essential for constructing the classical representation of each quantum state, i.e., $(\mathcal{C}^{(i)}, \bzi)$. In this regard, the computation cost amounts to the shot number $M$, as $\bzi$ can be acquired without access to quantum systems. Furthermore, as interpreted below, $M$ is modest in most QSL tasks. This characteristic \textit{sharply contrasts with many model-centric QSL proposals} \cite{ahmed2021quantum,zhang2021direct,lu2018separability}, which necessitate large or even an \textit{infinite number} of measurements to establish classical representations, thereby severely constraining their practical feasibility.

It is noteworthy that the computation cost under discussion does not consider the retrieval of data labels. This is because labeling annotations in QSL, analogous to various domains such as computer vision, are typically resource-intensive, and it is unreasonable to anticipate the efficient acquisition of labels. For this reason, we relax the condition of computational efficiency in label collection.

\noindent\textbf{For Criterion 3)}. Extensive theoretical studies have demonstrated the capacity of classical shadows to effectively tackle a diverse array of QSL tasks by using a modest shot number $M$, which typically exhibits polynomial scaling with respect to the qubit count $N$ \cite{elben2022randomized}. These results evidence that classical shadows convey ample information directly pertinent to the objectives of QSL, thereby forming a fundamental basis for the utilization of classical shadows in dataset construction to address a multitude of tasks.  Furthermore, the integration of readily accessible quantum system-specific information may enrich the capacity of learning models to distill knowledge from the dataset with a reduced $M$.  

\noindent \textbf{Remark}. The framework of data-centric QSL is not restricted to (Pauli-based) classical shadows. Alternative measurement protocols that meet the above three criteria could also serve as viable options.
 
\subsection{A general framework of data-centric QSL} 
The core insight driving the development of our dataset construction methodology is the belief that \begin{center}
 \textit{data-centric QSL can confer advantages upon diverse tasks, particularly those that classical shadows can effectively address}. 	
 \end{center}
 To illustrate this point more explicitly, we devise a general learning framework, dubbed ShadowNet, that leverages our proposed dataset to address different QSL tasks.

The scheme of ShadowNet is illustrated in Fig.~\ref{fig:scheme}(a). For a specified task, ShadowNet first retrieves a training dataset with $n$ examples, denoted by $\mathcal{D}_{\text{Tr}}=\{(\mathcal{C}^{(i)}, \bzi), \byi\}_{i=1}^n$, from the unified dataset $\mathcal{D}_{\text{QSL}}$. Once $\mathcal{D}_{\text{Tr}}$ is prepared, ShadowNet harnesses a tailored neural network to conduct the training. Note that the format of $(\mathcal{C}^{(i)}, \bzi)$ is not compatible with DNNs, necessitating a \textit{data feature engineering procedure} to transform the unprocessed data into a format amenable to DNN manipulation, i.e.,
\begin{equation}
	(\mathcal{C}^{(i)}, \bzi) \rightarrow \bxi.
\end{equation}
As elucidated in the subsequent two subsections, the strategy of data engineering is versatile, and the varied data feature engineering strategies may result in differing learning performances.  After the process of data feature engineering, ShadowNet optimizes the employed neural network to minimize the objective function  \begin{equation}\label{eqn:loss_shadownet}
 \Loss(\bW) = \frac{1}{n}\sum_{i=1}^n \loss\left(\mathcal{A}(\bxi;\bW), \byi \right), 
\end{equation} 
where $\mathcal{A}(\bx;\bW)$ denotes  the prediction of the neural network with $\bW$ being weights to be optimized and the per-sample loss $\loss(\bxi, \byi)$ quantifies the prediction error. The optimization of $\bW$ can be completed by gradient descent methods with a total of $T$ epochs.   

During the inference procedure, the optimized ShadowNet can efficiently predict unseen examples $\bx$, i.e., $\widetilde{\by}=\mathcal{A}(\bx, \bW^{(T)})$. A visualization of this procedure is shown in the lower panel of Fig.~\ref{fig:scheme}(a). Compared to prior NN-QSL methods and classical shadows, ShadowNet enables a reduced memory and computation cost to form $\bx$. In addition, a side benefit of employing classical shadows to form $\bx$ is quantifying the predictive faith of  $\widetilde{\by}$. As shown in Fig.~\ref{fig:scheme}(b), the prediction returned by ShadowNet should be located in the error bounds of its shadow estimation. Otherwise, the prediction is deemed to be \textit{unfaithful}, and the classical shadow estimation is preferred. Refer to Appendix A for a quantitative analysis of the predictive faith.

\begin{figure}[h!]
	\centering
\includegraphics[width=0.45\textwidth]{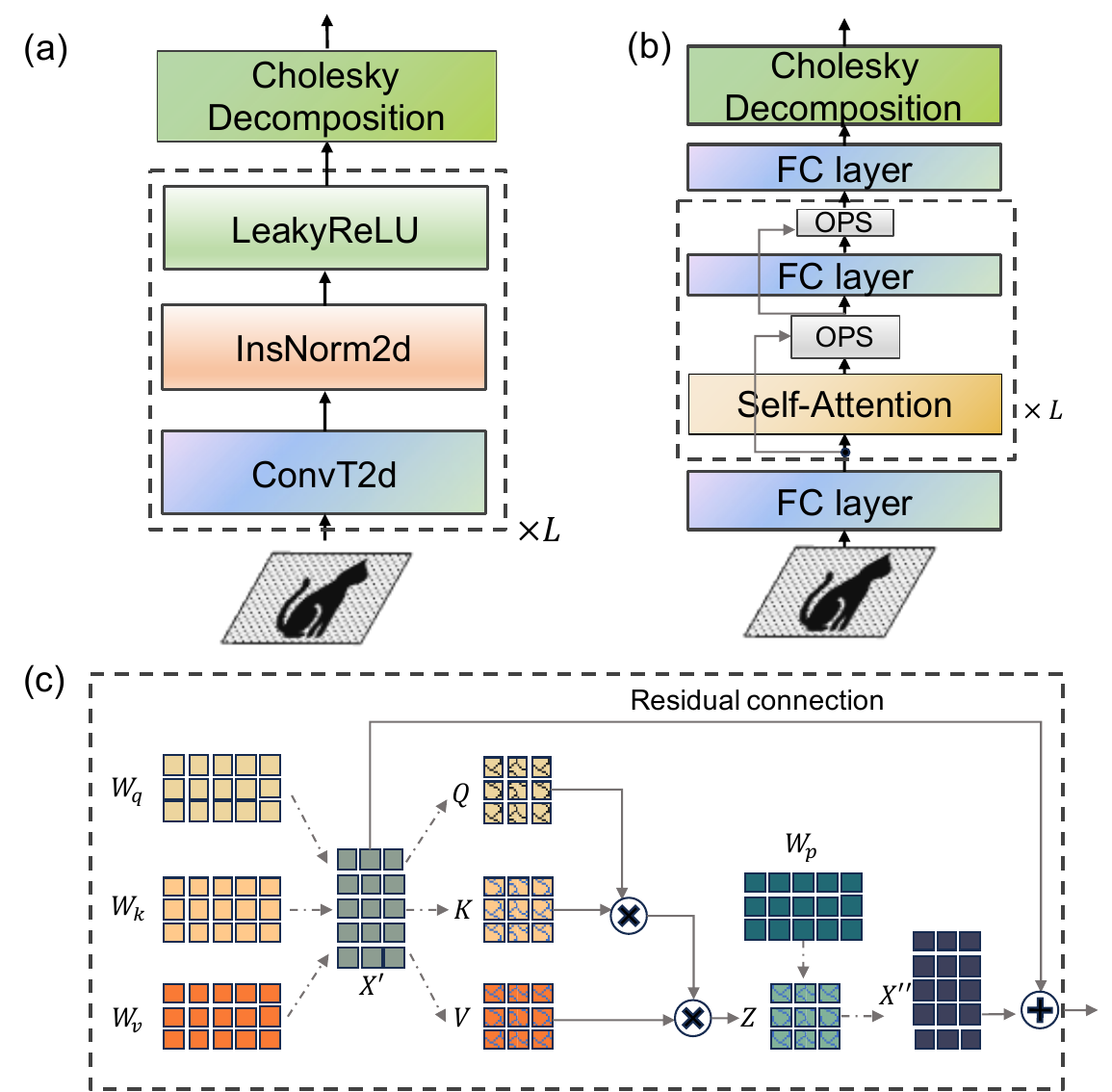}
\caption{\footnotesize{\textbf{Implementation of ShadowNet for QST}. Subplots (a) and (b) separately visualize the convolutional-based and attention-based ShadowNets when solving QST. The labels `ConvT2d', `Conv2d',  `InsNorm2d', and `LeakyReLU' refer to a two-dimensional transposed convolution operator, two-dimensional convolution operator, instance normalization layer, and LeakyReLU activation function, respectively. The labels `FC layer', `Self-attention', `OPS', and `Cholesky decomposition' refer to a fully-connected layer, self-attention block, standard operations in DNN (e.g., activation function and normalization layer), and  Cholesky decomposition layer, respectively. The symbol `$L$' stands for repeating the self-attention operation $L$ times.  Subplot (c) exhibits the implementation details of the self-attention layer adopted in the attention-based ShadowNet. }} 
\label{fig:append-CNN-shadowNet}
\end{figure}

\subsection{ShadowNet for quantum state tomography} 
Following the general framework of ShadowNet, several crucial components remain to be addressed to tackle quantum state tomography (QST). These include the approach to data feature engineering, the architecture of the employed neural network, and the specific formalism of the loss function. In the subsequent sections, we introduce details of these components when implementing the convolutional-based ShadowNet in Sec.~\ref{sec:cnn-shadowNet-QST} and the attention-based ShadowNet in Sec.~\ref{append:sec:implementation-shadowNet}.

\subsubsection{Convolutional-based ShadowNet in QST}\label{sec:cnn-shadowNet-QST}

\textbf{Data feature engineering}. For each example in $\mathcal{D}_{\text{Tr}}$, the procedure of  data feature engineering   adopted in the convolutional-based ShadowNet yields  
\begin{equation}\label{eqn:data-feature-shadow_QST-cnn}
	(\mathcal{C}^{(i)}, \bzi) \rightarrow \bxi = \left[\text{real}(  \hatrhoi), \text{img}(  \hatrhoi)\right], 
\end{equation} 
where $\bxi$ is a 3d array with the size $2^N\times 2^N \times 2$ that can be interpreted as an image with $2^N\times 2^N$ pixels and $2$ channels, $\hatrhoi$ is the reconstructed shadow state with  
\begin{equation}\label{eqn:shadow_QST}
	  \hatrhoi = \frac{1}{M}\sum_{m=1}^M \bigotimes_{j=1}^{N}  \Big(3U_{j,m}^{(i)\dagger} |b_{j,m}^{(i)}\rangle\langle b_{j,m}^{(i)}|U_{j,m}^{(i)} - \mathbb{I}_2\Big)\in \mathbb{C}^{2^N\times 2^N},
\end{equation}
and $\text{real}(\cdot)$ (or $\text{img}(\cdot)$) refers to the real (or imaginary) part of $\rhoi$. The operation of transforming complex numbers to real numbers facilitates the training of the handcrafted DNN by PyTorch.

\noindent\textbf{Model implementation}. The architecture of convolutional-based ShadowNet is exhibited in Fig.~\ref{fig:append-CNN-shadowNet}(a). The input data $\bxi$ is passed through a convolutional block $L$ times. Each block contains three components: a two-dimensional transposed convolution operator (ConvT2d) \cite{zeiler2010deconvolutional}, a two-dimensional Instance normalization layer (InsNorm2d) \cite{ulyanov2016instance}, and a LeakyReLU activation function. The role of ConvT2d is to upsample the input feature map instead of downsampling, which can increase the spatial resolution of an image. Here we adopt this operation to ensure the same dimension of the input and the final output. Moreover, InsNorm2d is employed to enhance trainability and stability. After passing $L$ convolutional blocks, the output interacts with the Cholesky decomposition layer, restricting the reconstructed state to be physical. 

\noindent\textbf{Objective function}.   The per-sample loss in Eq.~(\ref{eqn:loss_shadownet}) exploited to optimize the conventional-based ShadowNet is  
\begin{equation}\label{eqn:loss_QST}
	\loss(\bxi, \rhoi)= \|\tilderhoi - \rhoi\|_F^2,
\end{equation}
where  $\tilderhoi = \mathcal{A}(\bxi; \bW)$ denotes the reconstructed state of the $i$-th example output by the ShadowNet. 

\subsubsection{Attention-based ShadowNet in QST}\label{append:sec:implementation-shadowNet}

\textbf{Data feature engineering}. The data feature engineering procedure adopted in the attention-based ShadowNet is similar to the one adopted in the convolutional-based ShadowNet. The only modification is reshaping $\bxi$ in Eq.~(\ref{eqn:data-feature-shadow_QST-cnn}) to a two-dimensional array, i.e.,     
\begin{equation}\label{eqn:data-feature-att-QST}
	\bxi \xrightarrow{\text{Reshape}}  [\bx_s^{(i)}]_{s=1:R} \in\mathbb{R}^{2\times 4^N}.
\end{equation}
This reshaping operation aims to accord with the self-attention mechanism introduced in Eq.~(\ref{eqn:append-A-mechanism-attention}), where the data features $\bxi$ are segmented into $S=4^N$ pieces, where each piece $\bx_s$ has the dimension $p=2$.

\noindent\textbf{Model implementation}. The architecture of the attention-based ShadowNet is shown in  Fig.~\ref{fig:append-CNN-shadowNet}(b). Particularly, it first employs the fully-connected layer (i.e., labeled by the `FC' layer) to map each piece $\bx_s$ into a higher feature dimension. The mapped data, denoted by $X'$, are fed into a modified self-attention block proposed by Ref.~\cite{vaswani2017attention} $L$ times. Afterward, the output of the $L$-th modified self-attention block is operated with another fully-connected layer to project each piece into the original dimension $2$. Last, the projected feature interacts with the Cholesky decomposition layer in Sec.~\ref{sec:pre-cholesky}, restricting the reconstructed state to be physical.
  
 For self-consistency, we visualize the modified self-attention block in Fig.~\ref{fig:append-CNN-shadowNet}(c). Compared to the original mechanism introduced in Sec.~\ref{sec:layers-DNN-QSL}, the residual connection and layer normalization are involved. Concretely, the residual connection is highlighted by the marker `$\oplus$', summing the input $X'$ and the processed $X''$. The role of the weight matrix $W_p$ is to ensure the same dimension between $X'$ and $X''$. After summation, layer normalization is applied to normalize the distributions of intermediate layers, enabling smoother gradients, faster training, and better generalization accuracy \cite{xu2019understanding}. 
  
\noindent\textbf{Objective function}. The objective function is identical to the one used in convolutional-based ShadowNet.

\smallskip
\noindent\textbf{Remark}. We remark that, in the task of QST, both the convolutional-based ShadowNet and the attention-based ShadowNet can directly predict the density matrix of previously unseen states at the inference stage, using only the reconstructed shadow state as input, without requiring additional optimization. This feature pinpoints the efficiency of ShadowNet in comparison to optimization-based QST methods, which often entail lengthy post-processing procedures \cite{monz201114}. Besides, although fidelity is a standard metric for assessing the faithfulness in QST, exponential state copies are essential to ensure a reliable estimation. In this regard, a surrogate quantity $\mathsf{g}$ may be adopted to gauge the faithfulness, which leverages prior information about the explored quantum system. For example, when learning a class of ground states, $\mathsf{g}$ corresponds to the ground energies. The reconstructed state is judged to be faithful if the estimated ground state energy $\widetilde{\mathsf{g}}$ falls within the error bounds of its shadow estimation $\hat{\mathsf{g}}$.

\subsection{ShadowNet for direct fidelity estimation}
Direct fidelity estimation, in contrast to QST, typically pertains to scenarios involving a larger number of qubits. This distinction necessitates the novel data feature engineering approach, model implementation, and the objective function. For concreteness, in this subsection, we separately elucidate the realization of these components when the convolutional-based ShadowNet and the attention-based ShadowNet are employed.

\subsubsection{Convolutional-based ShadowNet in DFE}\label{append:subsec:ShadowNet-CNN}

\noindent\textbf{Data feature engineering}. Suppose that the noise of an $N$-qubit system can be described by a vector $\vec{p}$ with the dimension $d$. The procedure of data feature engineering applied to each example in $\mathcal{D}_{\text{Tr}}$ yields
\begin{equation}\label{eqn:att-DFE-data-reform}
(\mathcal{C}^{(i)}, \bzi) \rightarrow  	\bxi = 
\begin{bmatrix}
\text{real}(\hat{\rho}_1^{(i)}),  \text{img}(\hat{\rho}_1^{(i)}), \text{Dup}(\vec{p})\\
\cdots \\
\text{real}(\hat{\rho}_N^{(i)}), \text{img}(\hat{\rho}_N^{(i)}), \text{Dup}(\vec{p})
\end{bmatrix}^{\top},
\end{equation}
where $\bxi$ refers to a 2d array, $\hat{\rho}_j^{(i)}$ denotes the \textit{local inverse snapshot} at the $j$-th qubit with  \begin{equation}
\hat{\rho}_j^{(i)} = \frac{1}{M}\sum_{m=1}^M \left(3U_{j,m}^{(i)\dagger} |b_{j,m}^{(i)}\rangle\langle b_{j,m}^{(i)}|U_{j,m}^{(i)} - \mathbb{I}_2 \right),	
\end{equation}
and $\text{Dup}(\cdot)$ refers to the duplicating  operation to align the dimension of $\vec{p}$ with $\hat{\rho}_j^{(i)}$. To accommodate the architecture of the convolutional mechanism, $\bxi$ is reshaped to a three-dimensional array. For example, when $d=2$, the reshaped size could be $(N, 2, 10)$.   

\begin{figure}
	\centering
\includegraphics[width=0.49\textwidth]{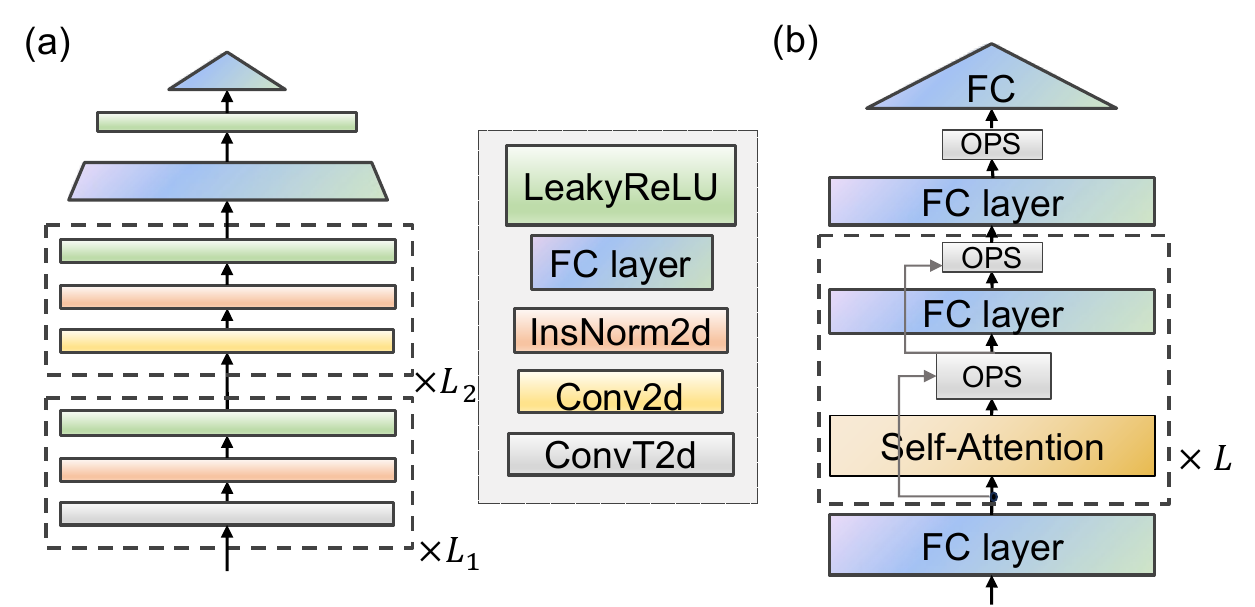}  
\caption{\footnotesize{\textbf{Implementation of ShadowNet for DFE}. Subplots (a) and (b) separately visualize the convolutional-based and attention-based ShadowNets when solving DFE. For the convolutional-based ShadowNet, the label `Conv2d' refers to a two-dimensional convolution operator, and the rest of the labels keep the same meaning as those in Fig.~\ref{fig:append-CNN-shadowNet}. The symbol `$L_1$' (or $L_2$ stands for repeating the operations in the dashed box $L_1$ (or $L_2$) times. For the attention-based ShadowNet, its layout is identical to the one introduced in Fig.~\ref{fig:append-CNN-shadowNet}, except for replacing the top layer with a fully connected layer. }} 
\label{fig:append-atten-dnn-visual}
\end{figure}

\noindent\textbf{Model implementation}. The architecture of the convolutional-based ShadowNet is exhibited in Fig.~\ref{fig:append-atten-dnn-visual}(a). Given access to the reformulated data, the convolutional-based ShadowNet applies $L_1$ convolutional blocks to upsample the input feature map. Afterwards, $L_2$  convolutional blocks are applied for downsampling. The only difference compared to the first $L_1$ blocks is replacing ConvT2d by the Convolution operator (Conv2d) to complete downsampling. Subsequently, two fully-connected layers accompanied by LeakyReLU activation function are applied to obtain the final prediction. 
 
 \noindent\textbf{Objective function}.  The per-sample loss in Eq.~(\ref{eqn:loss_shadownet}) yields   
\begin{equation}\label{eqn:loss-dfe}
\loss(\widetilde{\bm{y}}^{(i)}, \byi)= |\widetilde{\bm{y}}^{(i)} - \byi|^2,	
\end{equation}
which measures the estimation error of fidelity, and the corresponding label represents the accurate fidelity, i.e.,
\begin{equation}
	\byi=F_Q\left(\rhoi, \sigmai\right).
\end{equation}

\subsubsection{Attention-based ShadowNet in DFE}\label{sec:implementation-shadowNet-DFE}
\noindent\textbf{Data feature engineering}. Following the notations in Eq.~(\ref{eqn:att-DFE-data-reform}), the procedure of data feature engineering applied to each example in $\mathcal{D}_{\text{Tr}}$ yields
\begin{equation} 
(\mathcal{C}^{(i)}, \bzi) \rightarrow  	\bxi = 
\begin{bmatrix}
\text{real}(\text{vec}(\hat{\rho}_1^{(i)})), \text{img}(\text{vec}(\hat{\rho}_1^{(i)})), \vec{p}\\
\cdots \\
\text{real}(\text{vec}(\hat{\rho}_N^{(i)})), \text{img}(\text{vec}(\hat{\rho}_N^{(i)})), \vec{p}
\end{bmatrix}^{\top},
\end{equation}
where $\text{vec}(\cdot)$ refers to the vectorization operation. Then, for the purpose of adapting to the attention mechanism, the input data $\bxi$ is segmented into $N$ pieces, where each piece $\bxi_s$ in Eq.~(\ref{eqn:append-A-mechanism-attention}) has the dimension $p=8+d$, i.e., 
\begin{equation}
	\bx_s=\left[\text{real}(\text{vec}(\hat{\rho}_s^{(i)})), \text{img}(\text{vec}(\hat{\rho}_s^{(i)})),  \vec{p}\right]^{\top}.
\end{equation}

\noindent\textbf{Model implementation}. The implementation of attention-based ShadowNet is shown in  Fig.~\ref{fig:append-atten-dnn-visual}(b). Specifically, it applies the fully-connected layer to map each piece into a higher dimension, followed by applying $L$ modified attention blocks as aforementioned. Afterward, the output of the $L$-th modified self-attention block is operated with another fully-connected layer to project each piece into a lower dimension, associated with an average pooling operation to perform multi-scale summarization, as the prerequisite of the scalar prediction \cite{he2015spatial}. Last, a fully-connected layer is employed to output the final prediction. Denote $\widetilde{\bm{y}}^{(i)}$ as the prediction of the attention-based ShadowNet for the $i$-th example $\bxi$. 

 \noindent\textbf{Objective function}. The objective function follows Eq.~(\ref{eqn:loss-dfe}).

 \begin{figure*}
	\centering
\includegraphics[width=0.99\textwidth]{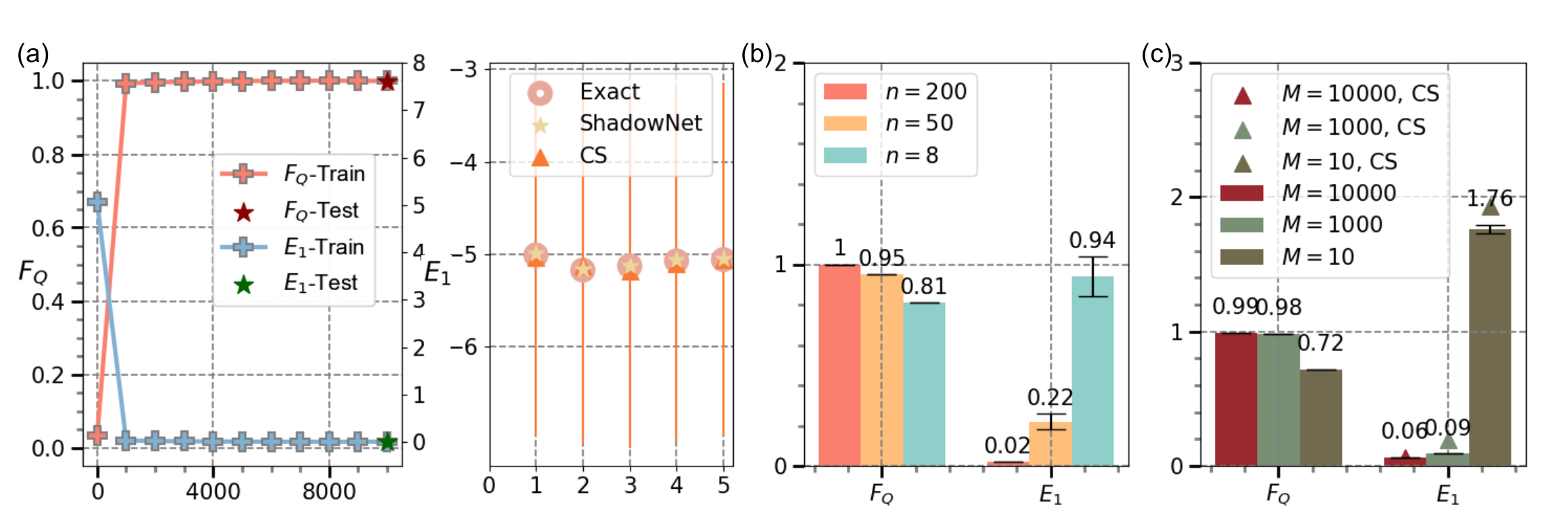}
	\caption{\footnotesize{\textbf{Results of ShadowNet in reconstructing ground states of TFIM with $N=5$}. (a) The left panel shows the train and test performance of ShadowNet with $n=200$ and $M=10^4$ with respect to epochs $T$. The notations `$F_Q$-Train' (`$F_Q$-Test') and `$\mathtt{E}_1$-Train' (`$\mathtt{E}_1$-Test') represent the fidelity between the averaged output of ShadowNet and the target state, and the estimation error on the training (test) dataset, respectively. The right panel illustrates the predictive faithfulness of ShadowNet on five test instances. The notation `CS' refers to classical shadows. The vertical line stands for the error bound of classical shadows. (b) Test performance of ShadowNet with respect to the varied size of the training dataset. The symbol $n=a$ refers that the size of $\mathcal{D}_{\text{Tr}}$ is $n$. The vertical bar highlights the variance of the collected results. (c) Test performance of ShadowNet with respect to the varied shot number $M$. The symbols share the same meaning as those in (a) and (b).}}
	\label{fig:sim-res-EShadow-Spin}
\end{figure*}

\noindent\textbf{Remark}. The proposed data feature engineering approaches embrace two compelling benefits: it enables the scalability of ShadowNet in handling large-scale qubit systems, as the dimension of $\bxi$ linearly scales with the qubit count $N$;  the incorporation of accessible information $\bzi$ into $\bxi$ allows ShadowNet to effectively capture the mapping rule connecting the measured data to precise fidelity values. Note that such approaches can be used for diverse QSL tasks beyond direct fidelity estimation, such as cross-platform fidelity estimation~\cite{qian2024multimodal}, entanglement entropy estimation~\cite{koutny2023deep}, and prediction of quantum phase transition~\cite{carleo2019machine}.

\section{Experiments}\label{sec:num-sim}
 
In this section, we conduct systematic simulations to explore the capability of ShadowNet in manipulating QST and DFE tasks. All numerical simulations are implemented using Julia, PennyLane, and PyTorch frameworks, and trained on four NVIDIA Tesla P40 GPUs. Furthermore, for all tasks, the AdamW optimizer is adopted to update the weights of ShadowNet. The hyperparameters related to AdamW are kept the same. That is, the initial learning rate is $0.0002$, and the coefficients used for computing running averages of the gradient and its square are 0.9 and 0.99.

\subsection{Quantum state tomography} 
We test ShadowNet on reconstructing ground states of two quantum spin systems that have been widely explored in many-body quantum simulations: the one-dimensional transverse-field Ising model (TFIM) and the XXZ model \cite{torlai2018neural,zhu2022flexible,carrasquilla2019reconstructing,schmale2022efficient}. The explicit expression of TFIM  is
\begin{equation}
	H_{\text{TFIM}} = J_z\sum_{\langle i,j \rangle}Z_iZ_j - J_x\sum_i X_i,
\end{equation}
where  $\langle i,j \rangle$ denotes the neighborhood spins, and $J_z$ and $J_x$ are two scalars standing for interaction strength and the transverse field, respectively. The phase of TFIM is ferromagnetic (antiferromagnetic) if $J_z <0$ ($J_z >0$).  The explicit form of XXZ model is 
\begin{equation}
	H_{\text{XXZ}} = -\sum_{i=0}^{N-2} \Delta_i (X_iX_{i+1} + Y_iY_{i+1}) - Z_iZ_{i+1},
\end{equation}
where $\Delta_i$ refers to the coupling parameter. For the XXZ model, it is in the ferromagnetic phase if $\Delta_i\in (-1, 1)$; otherwise, it is in the XY phase.  

In all simulations, the number of qubits is fixed to be $N=5$ for both Hamiltonians. To collect the training and test examples, we fix $J_x=1$ and sample $J_Z$ uniformly from a fixed interval specified below to obtain different examples. For XXZ model, we set $\Delta_i=\Delta_{i'}$ for $\forall i', i\in\{0,...,N-2\}$ and uniformly sample $\Delta_i$ from $[-3, 3]$ to generate training and test examples.  For each example,  the label (i.e., ground state) and the surrogate quantity (i.e., ground energy) are obtained by the exact diagonalization method, and the data feature (i.e., classical shadows) is obtained by PennyLane \cite{bergholm2018pennylane}. The performance of ShadowNet is evaluated by two metrics:  the fidelity $F_Q(\rho, \widetilde{\rho}) = (\Tr(\sqrt{\sqrt{\rho}\widetilde{\rho} \sqrt{\rho}}))^2$; the estimation error of the ground energy $\mathtt{E}_1=|\Tr((\widetilde{\rho}- \rho) H)|$. 

\noindent \textbf{Remark}. To highlight ShadowNet’s ability to extract useful information from input classical shadows $\hat{\rho}$ in order to infer the target ground state, we also report the ground state energy $\mathtt{E}_1$ directly estimated by $\hat{\rho}$ prior to being fed into ShadowNet. Since classical shadows are designed for estimating mean values, we deliberately avoid presenting the fidelity $F_Q(\rho, \hat{\rho})$ between the target ground state and the state reconstructed solely from classical shadows.

\smallskip
We first exploit the performance of the attention-based ShadowNet when applied to learn the ground states of TFIM from different perspectives. The implementation of the attention-based ShadowNet is as follows. According to Sec.~\ref{append:sec:implementation-shadowNet}, the attention block is fixed to be $L=3$. The first fully-connected layer maps the input feature from $2$ dimensions to $32$ dimensions. The dimension of the weight matrices $W_q$, $W_k$, and $W_v$, defined in Eq.~(\ref{eqn:append-A-mechanism-attention}), is set as $d_k=d_v=32$.   

\noindent\textbf{Learning dynamics}. We apply attention-based ShadowNet to learn the ground state reconstruction of TFIM. The number of measurements for classical shadows is $M=10000$. In addition, $J_Z$ is uniformly sampled from the interval $[-0.5, 0.5]$. When the size of the training dataset is $n=200$, the simulation results are exhibited in Fig.~\ref{fig:sim-res-EShadow-Spin}(a). The left panel indicates that after training $T=1000$ epochs, the fidelity  $F_Q$ over training examples is near to $1$ and the estimation error of the corresponding ground energy $\mathtt{E}_1$ is around zero. Moreover, ShadowNet has a satisfactory generalizable ability after $T=10000$ epochs. The test fidelity on $200$ unseen ground states is close to $1$. Meanwhile, the ground energy estimation error is almost zero, which is $0.02$. The right panel showcases the predictive faith of ShadowNet on five test instances in terms of $\mathtt{E}_1$,  reflecting that the predictions of the attention-based ShadowNet fall into the error bound of classical shadows and are closer to the exact results.

\noindent\textbf{Role of data size $n$}. We then explore how the dataset size  $n$ affects the performance of attention-based ShadowNet. To do so, all hyperparameter settings are kept the same as in the above experiment, except for the varied dataset size, i.e., $n\in\{8, 50, 200\}$. The simulation results are visualized in Fig.~\ref{fig:sim-res-EShadow-Spin}(b). When decreasing $n$ from $200$ to $8$, the fidelity over test examples drops from $1$ to $0.81$, and the test estimation error increases from $0.02$ to $0.94$. These results suggest that a modest-sized dataset is necessary to warrant the performance of ShadowNet. Moreover, the observed performance gap between ShadowNet and classical shadows in terms of $\mathtt{E}_1$ demonstrates that ShadowNet can effectively leverage limited measurement data to faithfully infer the target quantum state, further underscoring the potential of data-centric QSL.

\noindent\textbf{Role of shot number $M$}. The last task related to TFIM is investigating the role of the shot number $M$. All hyperparameter settings are the same as the previous experiments, except for two modifications. First, the shot number has three varied settings, i.e., $M\in \{10, 1000, 10000\}$. Second, when constructing training and test datasets, the range of the interaction strength $J_z$ expands to $[-2, 2]$, leading to increased diversity of examples. The simulation results of ShadowNet are shown in Fig.~\ref{fig:sim-res-EShadow-Spin}(c). An immediate observation is that when $M$ exceeds a threshold, the performance of ShadowNet tends to be optimal. In other words, the shot number $M$ should be carefully selected in ShadowNet, since continuously increasing $M$ not only require expensively computational cost, but also narrows the advantage of ShadowNet compared to classical shadows.

\smallskip
We then investigate the capability of attention-based ShadowNet when manipulating more difficult QST tasks. Precisely, the training dataset $\mathcal{D}_{\text{Tr}}$ contains the equivalent number of ground states from the TFIM and XXZ models. We fix $T=5000$, and vary the settings of $n$ and $M$ with $n\in \{100, 800\}$ and $M\in \{500, 1000, 2000\}$. The construction of ShadowNet is identical to the one used in prior tasks.

 \begin{table}[h]
\centering
\caption{\footnotesize{Inference performance of ShadowNet and classical shadows in reconstructing the ground states of TFIM and XXZ under varied settings. The notations $F_Q$, $\mathtt{E}_1$, and $\hat{\mathtt{E}}_1$ refer to the test fidelity of ShadowNet, and test energy estimation error by ShadowNet and classical shadows, respectively.}}
\label{tab:my-table}
\resizebox{.49\textwidth}{!}{%
\begin{tabular}{ccccc}
\hline
\multicolumn{1}{l}{} & \multicolumn{1}{l}{} & \multicolumn{3}{c}{Performance}                                       \\ \cline{3-5} 
\begin{tabular}[c]{@{}c@{}}Shots \\ M\end{tabular} &
  \begin{tabular}[c]{@{}c@{}}Size\\ n\end{tabular} &
  $F_Q$ &
  $\mathtt{E}_1$ &
  $\hat{\mathtt{E}}_1$ \\ \hline
\multirow{2}{*}{500} &
  100 &
  $0.878\pm 0.004$ &
  $1.361\pm 0.156$ &
  \multirow{2}{*}{\begin{tabular}[c]{@{}c@{}}0.477 \\ $\pm$ 0.204\end{tabular}} \\
                     & 800                  & $0.995\pm 0.204\times 10^{-8}$ & $0.044\pm 0.925\times 10^{-7}$    &  \\ \hline
\multirow{2}{*}{1000} &
  100 &
  $0.878\pm 0.352\times 10^{-3}$ &
  $1.342\pm 1.479$ &
  \multirow{2}{*}{\begin{tabular}[c]{@{}c@{}}0.308 \\ $\pm$ 0.102\end{tabular}} \\
                     & 800                  & $0.996\pm 0.438\times 10^{-8}$ & $0.042 \pm  0.360\times10^{-6} $ &  \\ \hline
\multirow{2}{*}{2000} &
  100 &
  $0.933\pm 0.304\times 10^{-3}$ &
  $0.488\pm 0.015 $ &
  \multirow{2}{*}{\begin{tabular}[c]{@{}c@{}}0.204 \\ $\pm$ 0.039\end{tabular}} \\
                     & 800                  & \bm{$0.996\pm 0.524\times 10^{-9}$} & \bm{$0.038 \pm 0.168\times10^{-6} $} &  \\ \hline
\end{tabular}%
}
\end{table}

\noindent\textbf{Performance of ShadowNet in learning both TFIM and XXZ}. The achieved results are summarized in Table \ref{tab:my-table}. Particularly, when $M\geq 500$ and $n=800$, the test fidelity is almost optimal, and the estimation error is near to zero. These observations accord with the results of TFIM, i.e., the size of the training dataset $n$ dominates the performance of ShadowNet when $M$ exceeds a threshold. In addition, ShadowNet dramatically outperforms classical shadows with sufficient training data, i.e., $0.044$ versus $0.477$ when  $n=800$ and $M=500$. This result underscores the strength of ShadowNet as a learning-based framework, capable of generalizing from limited training data to efficiently predict the properties of complex quantum systems with reduced sampling requirements.

 \begin{figure}
	\centering 
\includegraphics[width=0.48\textwidth]{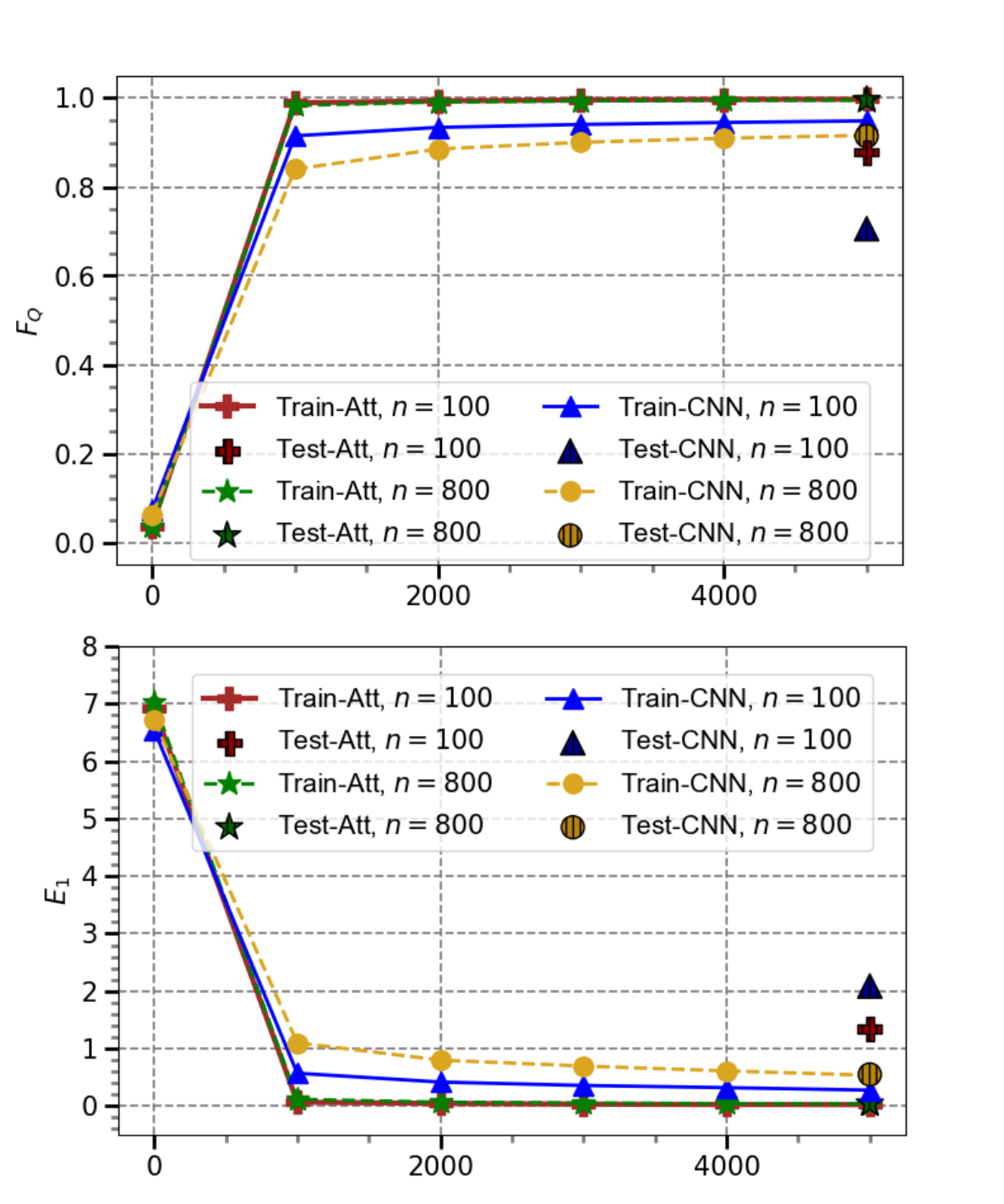}
\caption{\footnotesize{\textbf{Comparison of attention-based and convolutional-based ShadowNet in QST}. The upper and lower panels illustrate the performance of two types of  ShadowNet in the measure of fidelity and ground state energy estimation, respectively. The labels `Train-att, n=$a$' and `Test-att, n=$a$' refer to that when ShadowNet is implemented with the attention mechanism, with the number of training examples being $a$, the average fidelity (upper panel) or average energy estimation error (lower panel) of training examples and test examples, respectively. Following the same routine, the labels and `Train-CNN, $n=a$' and `Test-CNN, $n=a$' refer to the relevant results of ShadowNet implemented by the convolutional mechanism. The x-axis represents the number of epochs.  }}	\label{fig:append:sim-qst-spin}
\end{figure} 

\begin{figure*}
	\centering
\includegraphics[width=0.99\textwidth]{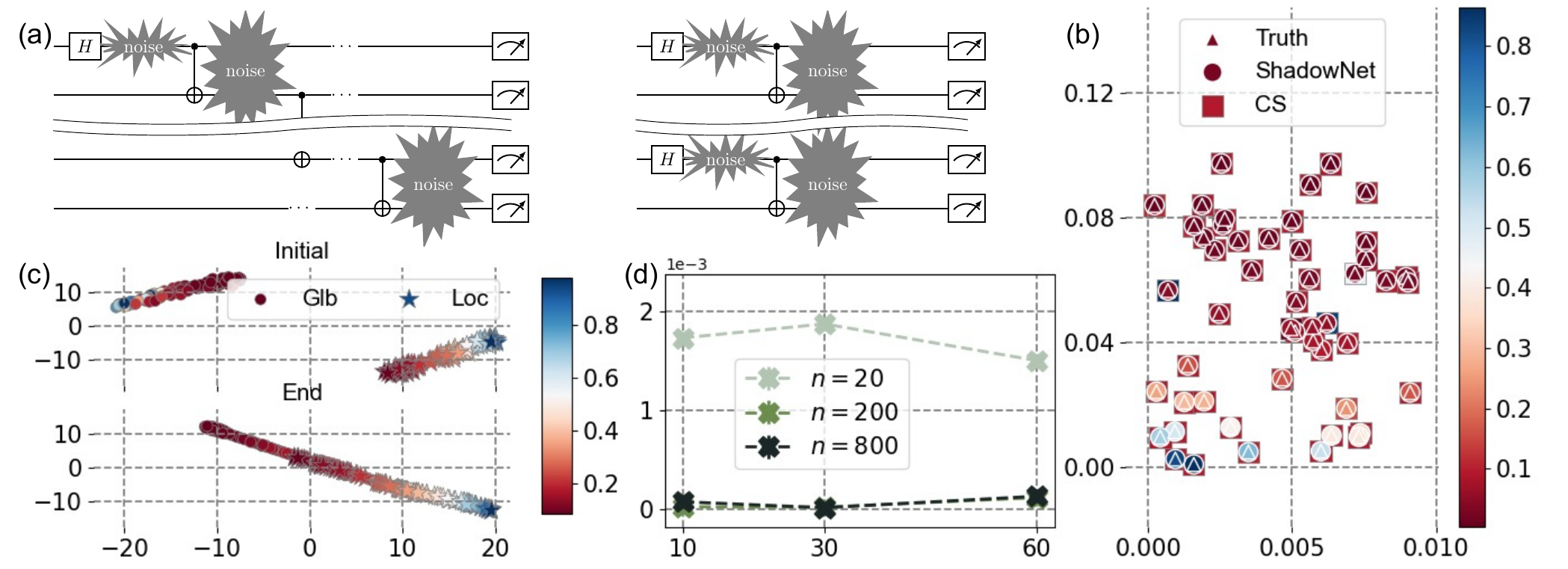}
	\caption{\footnotesize{\textbf{Results  of ShadowNet in solving DFE with noisy GHZ states}. (a) Circuit implementation in preparing global (left panel) and local (right panel) GHZ states under different levels of noise. (b) Performance of classical shadows (denoted by `CS') and ShadowNet in predicting $50$ noisy global GHZ states with $N=60$. The X-axis and Y-axis refer to the level of noise $p_1$ and $p_2$, respectively. The color bar stands for the fidelity between the prepared and ideal states. (c) Visualization of hidden features extracted by ShadowNet before and after training on the whole test examples. The label `Glb' (`Loc') represents the global (local) GHZ states. The color bar refers to the ground truth of fidelity of each state. (d) The scalability of ShadowNet trained on the varied training dataset size $n$ and the number of qubits $N$. The X-axis refers to the number of qubits and the Y-axis refers to the test loss. The label $n=a$ denotes that the number of training examples is $a$.  }}
	\label{fig:sim-res-EShadow-GHZ-global}
\end{figure*}

\smallskip
We last compare the performance of attention-based and convolutional-based ShadowNet in learning ground states of TFIM and XXZ models described above. The hyperparameter settings are as follows. We fix $M=1000$ and vary the number of training examples as $n=\{100, 800\}$. For TFIM, we sample $J_Z$ uniformly from the interval $[-0.5, 0.5]$ to obtain different examples. The implementation of attention-based ShadowNet is the same as the one used in previous tasks. The implementation of the convolutional-based ShadowNet is as follows. According to Sec.~\ref{sec:cnn-shadowNet-QST}. The number of encoding blocks is set as $L=5$. The kernel size, stride, and padding of ConvT2d in each block are identical, which is equal to $3$, $1$, and $1$, respectively. The only difference among convolutional blocks is the number of channels, which is $2\rightarrow 256 \rightarrow 192 \rightarrow 192 \rightarrow 64 \rightarrow 32 \rightarrow 2$.   We repeat the simulation with five times under different random seeds to collect the statistical results.

\noindent\textbf{Comparison of attention-based and convolutional-based ShadowNet}. The simulation results are shown in Fig.~\ref{fig:append:sim-qst-spin}. The upper panel depicts the performance of ShadowNet implemented by the attention mechanism and the convolution mechanism in terms of fidelity. After $4000$ epochs, the averaged fidelity over training examples is near to $F_Q=1$ for the attention mechanism, in both cases of $n=100$ and $n=800$. However, the averaged fidelity over training examples under the convolutional mechanism is imperfect, which is $F_Q= 0.706$ with $n=100$ and $F_Q= 0.914$ with $n=800$. In addition, for both attention and convolutional mechanisms, ShadowNet experiences the overfitting phenomenon when $n=100$. By increasing the training examples to $n=800$, this issue can be addressed, in which the difference in fidelity between training examples and test examples is ignorable. The lower panel shows the performance of ShadowNet implemented by the attention mechanism and the convolution mechanism in the measure of the estimation error of the ground state energy. Among different settings, only ShadowNet implemented by the attention mechanism with $n=800$ attains the ignorable estimation error over the test examples.

The distinct learning performance provides the following insights. First, the performance may be affected by the employed architecture of the DNN. The inferior performance of the convolutional mechanism compared to the attention mechanism may be caused by its insufficient ability to capture the long-range correlation of data features. In other words, a handcrafted architecture of DNN enables the improved learning efficiency of ShadowNet. Second, the number of training examples plays an important role in the performance of ShadowNet. Insufficient training examples may incur overfitting and unsatisfactory performance of ShadowNet at the inference stage.

\subsection{Direct fidelity estimation}  
We examine the efficacy of ShadowNet towards DFE tasks by applying it to learn two classes of noisy GHZ states simultaneously. As pictured in Fig.~\ref{fig:sim-res-EShadow-GHZ-global}(a), the first class of states is the noisy global GHZ states, i.e., 
\begin{equation}
	\rho(\vec{p})=\mathcal{E}_{\vec{p}}(	\ket{\GHZ}),
\end{equation}
where $\mathcal{E}_{\vec{p}}(\cdot)$ is the local depolarization channel applied to the GHZ state $\ket{\GHZ} = {({\ket{0}^{\otimes N} + \ket{1}^{\otimes N}})}/{\sqrt{2}}$, and $\vec{p}=[p_1, p_2]$ describes the depolarization error rate of the single-qubit and two-qubit gates, respectively. The second class of states is the noisy local GHZ states, i.e.,
\begin{equation}
	\kappa(\vec{p})=\mathcal{E}_{\vec{p}}(	\ket{\GHZ_l}),
\end{equation}
where $\ket{\GHZ_l} = ({(\ket{0} + \ket{1})}/{\sqrt{2}})^{\otimes N}$. The label of each example amounts to the fidelity between the noisy and ideal states, i.e.,
\begin{equation}
	F_Q(\rho(\vec{p}),\ket{\GHZ})~\text{or}~ F_Q(\kappa(\vec{p}),\ket{\GHZ_l}).
\end{equation}
When collecting training and test examples, we uniformly sample $p_1\in [0.0001,0.01]$ and $p_2\in [0.0001,0.1]$ and fix the shot number for (Pauli-based) classical shadows to be $M=2000$. For each example, its classical shadows in the noisy scenario and the corresponding labels are collected by PastaQ \cite{pastaq}, a Julia software toolbox for the simulation of large-qubit quantum circuits. The learning performance and predictive faith are measured by  $\mathtt{E}_2= \mathsf{L}= |\widetilde{\bm{y}}-F_Q|^2$. Each setting is repeated five times to obtain the statistical results.

\smallskip
We first explore the capability of attention-based ShadowNet, whose implementation remains the same for all DFE tasks. Specifically, following Sec.~\ref{sec:implementation-shadowNet-DFE}, the first fully-connected layer maps the input feature from $10$ dimensions to $32$ dimensions, and the number of attention blocks is fixed to be $L=3$. The dimension of the weight matrices $W_q$, $W_k$, and $W_v$, defined in Eq.~(\ref{eqn:append-A-mechanism-attention}), is set as $d_k=d_v=32$. When the input data passes through three attention blocks, the corresponding feature interacts with the fully-connected layer with the hidden dimension $32$, followed by the adaptive average pooling and another fully-connected layer with the output dimension $1$.

\noindent\textbf{Comparison with classical shadows}. Our first task is comparing the accuracy of the fidelity estimation between classical shadows (i.e., $M=2000$) and ShadowNet when $N=60$ and $n=800$.  In particular, Fig.~\ref{fig:sim-res-EShadow-GHZ-global} (b) illustrates the predictive fidelity on $50$ test examples of noisy global GHZ states. ShadowNet can accurately predict the true fidelity, while classical shadows fail to compute the true fidelity.  This observation implies the separable capability between NN-QSL models and classical shadows in handling large-scale quantum systems. The inferior performance of classical shadows stems from the limited number of snapshots $M$. As proved in Appendix A, achieving a satisfactory estimation would require $M\sim \Theta({3^{60}})$ under Pauli-based classical shadows. To uncover the advantage of ShadowNet, we further adopt the t-distributed stochastic neighbor embedding (t-SNE) algorithm to visualize its learning dynamics during the training \cite{van2008visualizing}. Fig.~\ref{fig:sim-res-EShadow-GHZ-global}(c) plots the representations at the last-second layer of ShadowNet when projecting to a two-dimensional plane. Before training, noisy global and local GHZ states are located into two clusters. Meanwhile, the representations of noisy global GHZ states are mixed, implying that ShadowNet cannot accurately predict the fidelity of each state. After training, the representations of noisy global and local GHZ states are well-aligned, enabling accurate predictions.

\noindent\textbf{Role of data size $n$}. We next systematically comprehend how the size of the training dataset $n$ affects the prediction accuracy of the attention-based ShadowNet with respect to the increased number of qubits $N$. To do so, we vary the number of qubits as $N\in \{10, 30, 60\}$ and the dataset size as $n\in \{20, 200, 800\}$. The simulation results are demonstrated in Fig.~\ref{fig:sim-res-EShadow-GHZ-global}(d). When $n$ exceeds a threshold, e.g., $n>20$, ShadowNet attains a satisfactory performance for all settings of $N$. This phenomenon suggests that the improvement of ShadowNet over classical shadows becomes increasingly pronounced as the system size $N$ increases. In other words, our dataset construction rule has the potential to substantially reduce the required number of training examples, as a crucial characteristic to apply ShadowNet to solve large-scale DFE tasks.  

\noindent\textbf{Complementary role of classical shadows and noise parameter}. Recall that the data features of ShadowNet incorporate classical shadows and the noisy parameter in solving DFE tasks. A natural question is understanding whether ShadowNet can effectively learn the mapping rule only using noise parameters or classical shadows as the data features. To address this issue, we conduct simulations with two modified datasets. In the first modified dataset, we restricted the data features to include only the noise parameters (i.e., the dimension of each example reduces from $N\times 10$ to $N\times 2$), while in the second modified dataset, we included only classical shadows in the data features of each example (i.e., the dimension of each example reduces from $N\times 10$ to $N\times 8$). Note that apart from the modifications in the datasets, the model implementation of the attention-based ShadowNet remains unchanged. Moreover, the hyperparameter settings are almost the same as in prior tasks, i.e., the number of qubits is $N=60$, the number of measurements to collect classical shadows is $M=2000$, and the size of training examples is $n=600$.

\begin{figure}
	\centering
\includegraphics[width=0.48\textwidth]{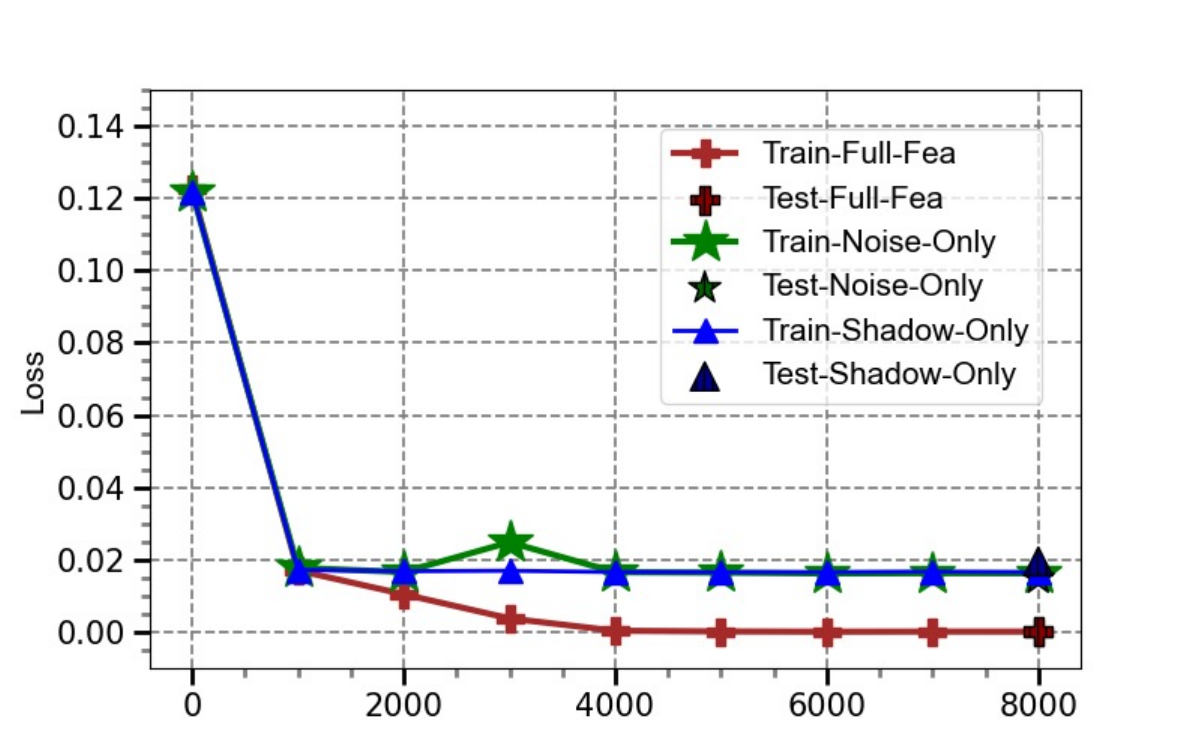}
\caption{\footnotesize{\textbf{The role of classical shadows and noise parameters in data features}. The plot depicts the dynamics of loss during the training procedure.  The labels `Train-full-Fea' and `Test-full-Fea' refer to the loss of ShadowNet over training and test examples using full data features, respectively. Similarly, `Train-Noise-Only' (`Train-Shadow-Only') and `Test-Noise-Only' (`Test-Shadow-Only') refer to the loss of ShadowNet over training and test examples only using noise parameters (classical shadows) as data features, respectively. }}
\label{fig:append:DFE-ablation} 
\end{figure}

The simulation results are presented in Fig~\ref{fig:append:DFE-ablation}. While the training loss consistently decreases during optimization for all datasets, we observe that the attention-based ShadowNet trained on the dataset with full features notably outperforms the other two settings. This advantage is preserved when evaluating the optimized ShadowNet on the test examples, where the average test loss is $0.00013$ for the full feature dataset, $0.0147$ for the dataset containing only noise parameters, and $0.0196$ for the dataset containing only classical shadows. These results pinpoint the complementary role of noise parameters and classical shadows in using ShadowNet to complete DFE tasks.

\smallskip
We next compare the performance of the attention-based ShadowNet with the convolutional-based ShadowNet, where the relevant results can provide insights about how the architecture of DNN influences their power in DFE tasks. The implementation of convolutional-based ShadowNet follows the instructions in Sec.~\ref{append:subsec:ShadowNet-CNN}. Namely, we set $L_1=L_2=2$. From the first to the fourth convolutional layer, the kernel size is $3$, $(3,7)$, $(5,1)$, and $3$; the stride is $1$, $1$, $(2,1)$, and $2$; the padding is $1$, $1$, $0$, and $1$. The number of channels follows $2\rightarrow 256 \rightarrow 192 \rightarrow 16$. When the input passes through convolutional blocks, it interacts with two fully-connected layers with the dimensions $320$ and $128$, respectively, followed by a Sigmoid operation.   

\begin{figure}
	\centering
\includegraphics[width=0.45\textwidth]{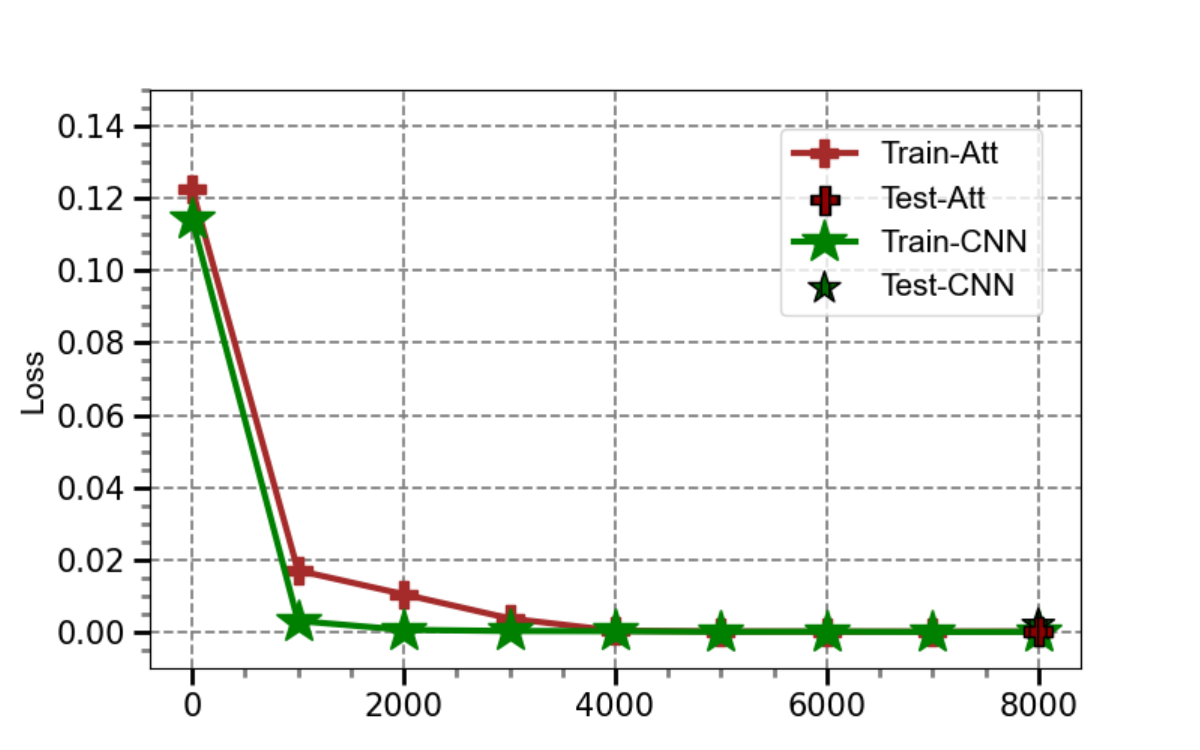}
\caption{\footnotesize{\textbf{Comparison of ShadowNet implemented by different mechanisms in solving the DFE task}. The plot depicts the loss dynamics of ShadowNet implemented by the attention and convolutional mechanisms. The labels `Train-Att' (`Train-CNN') and `Test-Att' (`Test-CNN') represent the performance of ShadowNet implemented by the attention (convolutional) mechanism over training and test examples, respectively. The $x$-axis stands for the number of epochs.   }}
\label{fig:append:ShadowNet-CNN-Att-DFE}
\end{figure}

\noindent\textbf{Comparison of attention-based and convolutional-based ShadowNet}.   To make a fair comparison, the convolutional-based  ShadowNet adopts the same hyper-parameter settings and optimization methods as those employed to train the attention-based ShadowNet, i.e., $N=60$, $n=800$, and $M=2000$. The simulation is repeated $5$ times to collect the statistical results.  The simulation results are shown in Fig.~\ref{fig:append:ShadowNet-CNN-Att-DFE}. For both mechanisms, the training and test losses are below $0.002$ after $8000$ epochs. The achieved results suggest that ShadowNet with the attention and convolutional mechanism can capture the mapping rule from the local shadows and noise parameters to the fidelity between noisy and ideal GHZ states. Moreover, combining the results in Fig.~\ref{fig:append:DFE-ablation}, we can conclude that the formalism of data features plays an important role in DFE tasks.

\begin{figure}
	\centering
\includegraphics[width=0.48\textwidth]{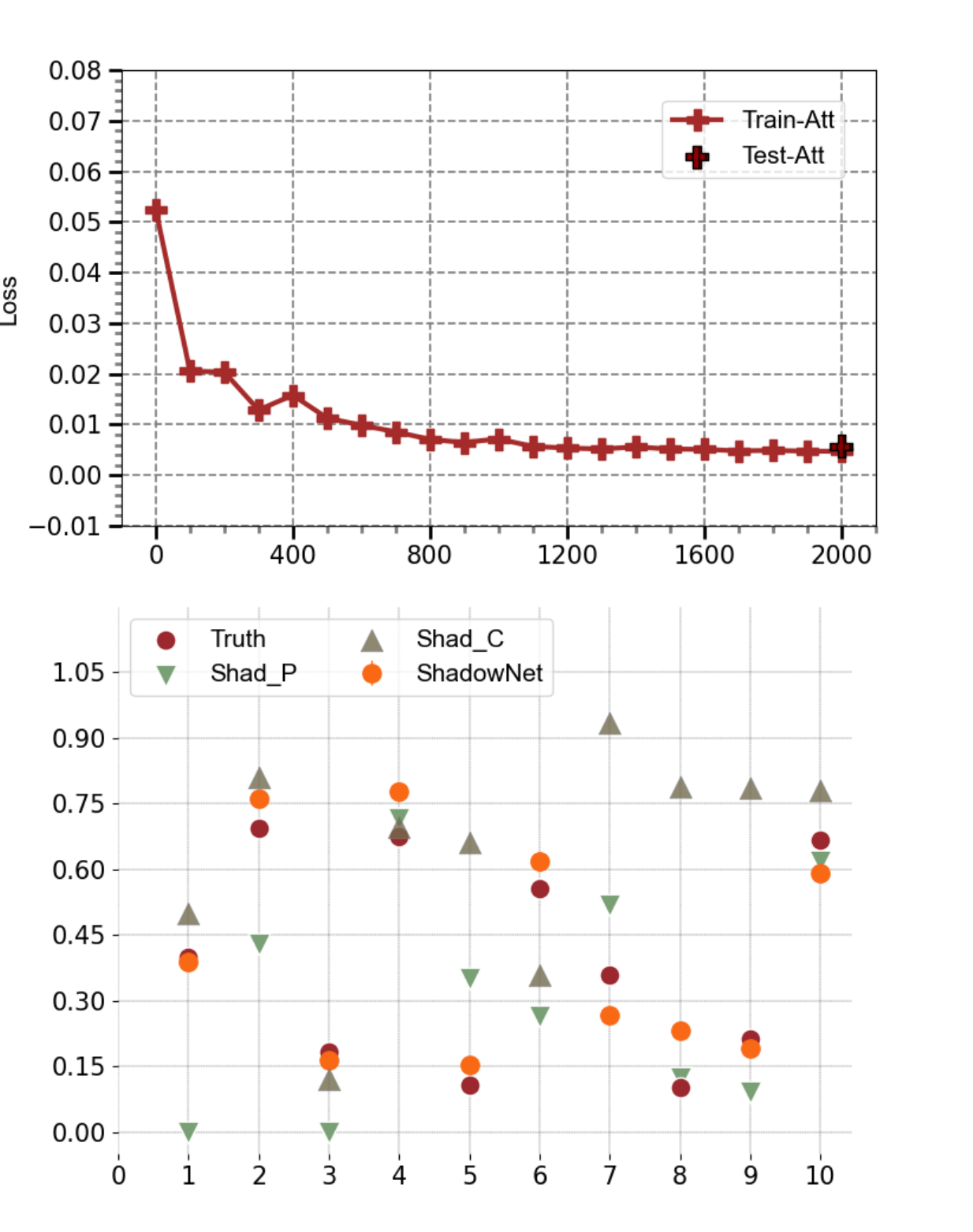}
\caption{\footnotesize{\textbf{Comparison of classical shadows of Pauli-based measurements and Clifford-based measurements, and ShadowNet}. The upper panel demonstrates the training and test loss of ShadowNet. The labels share the same meaning as those in Fig.~\ref{fig:append:ShadowNet-CNN-Att-DFE}. The lower panel shows the fidelity estimation of classical shadows and ShadowNet over $10$ test instances. The x-axis and y-axis separately represent the instance index and the estimated fidelity, respectively. The labels `Shad\_C' and `Shad\_P' refer to the estimated fidelity by classical shadows under the Clifford measurements and Pauli measurements, respectively. }}
\label{fig:append_DFE_Clifford}
\end{figure}

\smallskip
We lastly compare the predictive faithfulness of ShadowNet when classical shadows are completed by Clifford-based random measurements and Pauli-based random measurements. To do so, we conduct similar simulation tasks proposed by Ref.~\cite{huang2020predicting}, which focuses on the state preparation error rather than the local depolarization error considered before. In this new task, the class of noisy GHZ states yields 
\[\rho_p = (1-p)\ket{\GHZ}\bra{\GHZ} + p\ket{\perp}\bra{\perp}, \]
 where $p$ is the error rate and $\ket{\perp}$ refers to the orthogonal state and is set as $(\mathbb{I}_1\otimes X \otimes \mathbb{I}_{2^{N-2}}) \ket{0}^{\otimes N}$. Let the ideal state be an $N$-qubit $\ket{\GHZ}$. To collect the training and test examples, we randomly sample $p$ from $0.1$ to $0.9$. The noisy state $\rho_p$ with the varied $p$ is randomly measured in the Pauli basis $M$ times to form the local shadows as data features. The corresponding label equals $1-p$, which measures the fidelity between $\rho_p$ and $\ket{\GHZ}$. 
 
 The hyper-parameter settings are as follows.  We set the number of training examples as $n=580$, the number of qubits $N=4$, and the shot number is $M =100$. The total number of epochs is $T=2000$. The architecture of ShadowNet follows the aforementioned attention mechanism. The only modification is the input data feature, where the noise parameter is removed. This is because in this task, the target label corresponds to the noise parameter $p$. For each setting, we repeat the simulations five times to collect the statistical results. 

\noindent\textbf{Simulation results under Clifford measurement}.  The simulation results are shown in Fig.~\ref{fig:append_DFE_Clifford}. The training loss after $T$ epochs converges to $0.0045$. The discrepancy between the training and test losses is ignorable, implying the generalization ability of ShadowNet. We next employ the trained ShadowNet to understand the aforementioned predictive faithfulness issue. Particularly, we visualize the estimated fidelity of ShadowNet, classical shadows under Clifford and Pauli measurements with $500$ shots, and ground truth over $10$ test instances in the lower panel of Fig.~\ref{fig:append_DFE_Clifford}.  In this setting, the estimation error of classical shadows under the Clifford measurements is upper-bounded by $0.63$ with probability at least $0.8$ (see Appendix A for details). Despite the large error bound, an interesting phenomenon is that the estimated fidelity generally remains proximal to the ground truth. In other words, a modest number of snapshots allows classical shadows with Clifford measurements to provide a good predictive faith in the average case. Besides, for most instances, ShadowNet outperforms classical shadows with the varied measurement groups, suggesting the advantage of ShadowNet. 
  
\noindent \textbf{Remark}. For completeness, we append an experiment to examine the capability of ShadowNet for nonlinear-property prediction. Specifically, we use the same GHZ-state setting and take purity prediction as the target task. Refer to Appendix~B for details.

\section{Discussion and outlook}\label{sec:discussion}
We present the concept of data-centric QSL, emphasizing the central role played by datasets in contrast to the model-centric approaches prevalent in previous NN-QSL methods, which primarily focus on the design of DNN architectures. Central to data-centric QSL is the utilization of classical shadows, known for their memory efficiency and estimation guarantees, to create datasets. In this respect, we devise ShadowNet as a flexible and robust paradigm to tackle QSL tasks, with a specific emphasis on the synergy of generalizability, efficiency, and faithfulness.  Due to the flexibility of ShadowNet, we devise two distinct approaches for dataset construction and model implementation, tailored to the specific requirements of the task at hand. Numerical simulations confirm the efficacy of ShadowNet and exhibit its potential to advance data-centric QSL.

Our work stimulates several important avenues for future research. While our focus has been on Pauli-based random measurements due to their hardware-friendly feature, exploring advanced variants of classical shadows presents an intriguing opportunity to enhance the performance of ShadowNet \cite{nguyen2022optimizing,huang2022learning,bertoni2022shallow,hadfield2022measurements,struchalin2021experimental,wei2023neural}. Moreover, instead of the discrete quantum systems, the applicability of ShadowNet can be extended to facilitate the learning of continuous-variable quantum systems by leveraging continuous-variable classical shadows to construct the similar training datasets \cite{becker2022classical}.

Exploring alternative construction rules and expanding the applications of ShadowNet represent a critical avenue for further research. In the context of DFE, we leverage the noise information inherent in quantum systems as part of data features. As the majority of QSL problems arise from noisy intermediate-scale quantum devices \cite{preskill2018quantum}, it is intriguing to investigate whether other easily accessible information pertaining to quantum devices can be utilized to construct datasets that aid DNN in extracting underlying mapping rules, even with limited data sizes. For instance, considering the prominence of variational quantum algorithms \cite{cerezo2021variational,bharti2021noisy,tian2023recent,li2022recent} as a major application for near-term quantum devices, the design of novel data-centric QSL models to enhance the characterization and error mitigation of the output of these algorithms \cite{jnane2023quantum,bennewitz2022neural}  emerges as a desirable direction.


\begin{thebibliography}{10}

\bibitem{eisert2020quantum}
Jens Eisert, Dominik Hangleiter, Nathan Walk, Ingo Roth, Damian Markham, Rhea
  Parekh, Ulysse Chabaud, and Elham Kashefi.
\newblock Quantum certification and benchmarking.
\newblock {\em Nature Reviews Physics}, 2(7):382--390, 2020.

\bibitem{gebhart2023learning}
Valentin Gebhart, Raffaele Santagati, Antonio~Andrea Gentile, Erik~M Gauger,
  David Craig, Natalia Ares, Leonardo Banchi, Florian Marquardt, Luca
  Pezz{\`e}, and Cristian Bonato.
\newblock Learning quantum systems.
\newblock {\em Nature Reviews Physics}, 5(3):141--156, 2023.

\bibitem{gross2010quantum}
David Gross, Yi-Kai Liu, Steven~T Flammia, Stephen Becker, and Jens Eisert.
\newblock Quantum state tomography via compressed sensing.
\newblock {\em Physical Review Letters}, 105(15):150401, 2010.

\bibitem{blume2010optimal}
Robin Blume-Kohout.
\newblock Optimal, reliable estimation of quantum states.
\newblock {\em New Journal of Physics}, 12(4):043034, 2010.

\bibitem{flammia2012quantum}
Steven~T Flammia, David Gross, Yi-Kai Liu, and Jens Eisert.
\newblock Quantum tomography via compressed sensing: Error bounds, sample
  complexity and efficient estimators.
\newblock {\em New Journal of Physics}, 14(9):095022, 2012.

\bibitem{sugiyama2013precision}
Takanori Sugiyama, Peter~S Turner, and Mio Murao.
\newblock Precision-guaranteed quantum tomography.
\newblock {\em Physical Review Letters}, 111(16):160406, 2013.

\bibitem{haah2016sample}
Jeongwan Haah, Aram~W Harrow, Zhengfeng Ji, Xiaodi Wu, and Nengkun Yu.
\newblock Sample-optimal tomography of quantum states.
\newblock In {\em Proceedings of the Forty-Eighth Annual ACM Symposium on
  Theory of Computing}, pages 913--925, 2016.

\bibitem{o2016efficient}
Ryan O'Donnell and John Wright.
\newblock Efficient quantum tomography.
\newblock In {\em Proceedings of the Forty-Eighth Annual ACM Symposium on
  Theory of Computing}, pages 899--912, 2016.

\bibitem{kueng2017low}
Richard Kueng, Holger Rauhut, and Ulrich Terstiege.
\newblock Low rank matrix recovery from rank one measurements.
\newblock {\em Applied and Computational Harmonic Analysis}, 42(1):88--116,
  2017.

\bibitem{cramer2010efficient}
Marcus Cramer, Martin~B Plenio, Steven~T Flammia, Rolando Somma, David Gross,
  Stephen~D Bartlett, Olivier Landon-Cardinal, David Poulin, and Yi-Kai Liu.
\newblock Efficient quantum state tomography.
\newblock {\em Nature Communications}, 1(1):149, 2010.

\bibitem{landon2012practical}
Olivier Landon-Cardinal and David Poulin.
\newblock Practical learning method for multi-scale entangled states.
\newblock {\em New Journal of Physics}, 14(8):085004, 2012.

\bibitem{lanyon2017efficient}
BP~Lanyon, C~Maier, Milan Holz{\"a}pfel, Tillmann Baumgratz, C~Hempel,
  P~Jurcevic, Ish Dhand, AS~Buyskikh, AJ~Daley, Marcus Cramer, et~al.
\newblock Efficient tomography of a quantum many-body system.
\newblock {\em Nature Physics}, 13(12):1158--1162, 2017.

\bibitem{knill2008randomized}
Emanuel Knill, Dietrich Leibfried, Rolf Reichle, Joe Britton, R~Brad Blakestad,
  John~D Jost, Chris Langer, Roee Ozeri, Signe Seidelin, and David~J Wineland.
\newblock Randomized benchmarking of quantum gates.
\newblock {\em Physical Review A}, 77(1):012307, 2008.

\bibitem{magesan2012efficient}
Easwar Magesan, Jay~M Gambetta, Blake~R Johnson, Colm~A Ryan, Jerry~M Chow,
  Seth~T Merkel, Marcus~P Da~Silva, George~A Keefe, Mary~B Rothwell, Thomas~A
  Ohki, et~al.
\newblock Efficient measurement of quantum gate error by interleaved randomized
  benchmarking.
\newblock {\em Physical Review Letters}, 109(8):080505, 2012.

\bibitem{helsen2022general}
Jonas Helsen, Ingo Roth, Emilio Onorati, Albert~H Werner, and Jens Eisert.
\newblock General framework for randomized benchmarking.
\newblock {\em PRX Quantum}, 3(2):020357, 2022.

\bibitem{flammia2011direct}
Steven~T Flammia and Yi-Kai Liu.
\newblock Direct fidelity estimation from few {Pauli} measurements.
\newblock {\em Physical Review Letters}, 106(23):230501, 2011.

\bibitem{huang2020predicting}
Hsin-Yuan Huang, Richard Kueng, and John Preskill.
\newblock Predicting many properties of a quantum system from very few
  measurements.
\newblock {\em Nature Physics}, 16(10):1050--1057, 2020.

\bibitem{elben2022randomized}
Andreas Elben, Steven~T Flammia, Hsin-Yuan Huang, Richard Kueng, John Preskill,
  Beno{\^\i}t Vermersch, and Peter Zoller.
\newblock The randomized measurement toolbox.
\newblock {\em Nature Reviews Physics}, pages 1--16, 2022.

\bibitem{granade2012robust}
Christopher~E Granade, Christopher Ferrie, Nathan Wiebe, and David~G Cory.
\newblock Robust online {Hamiltonian} learning.
\newblock {\em New Journal of Physics}, 14(10):103013, 2012.

\bibitem{bairey2019learning}
Eyal Bairey, Itai Arad, and Netanel~H Lindner.
\newblock Learning a local {Hamiltonian} from local measurements.
\newblock {\em Physical Review Letters}, 122(2):020504, 2019.

\bibitem{huang2023learning}
Hsin-Yuan Huang, Yu~Tong, Di~Fang, and Yuan Su.
\newblock Learning many-body {Hamiltonians} with {Heisenberg}-limited scaling.
\newblock {\em Physical Review Letters}, 130(20):200403, 2023.

\bibitem{mayers2003self}
Dominic Mayers and Andrew Yao.
\newblock Self testing quantum apparatus.
\newblock {\em Quantum Information \& Computation}, 4(4):273--286, 2004.

\bibitem{montanaro2013survey}
Ashley Montanaro and Ronald De~Wolf.
\newblock A survey of quantum property testing.
\newblock {\em Theory of Computing}, pages 1--81, 2016.

\bibitem{vsupic2020self}
Ivan {\v{S}}upi{\'c} and Joseph Bowles.
\newblock Self-testing of quantum systems: A review.
\newblock {\em Quantum}, 4:337, 2020.

\bibitem{aaronson2018online}
Scott Aaronson, Xinyi Chen, Elad Hazan, Satyen Kale, and Ashwin Nayak.
\newblock Online learning of quantum states.
\newblock {\em Advances in Neural Information Processing Systems}, 31, 2018.

\bibitem{chen2022adaptive}
Xinyi Chen, Elad Hazan, Tongyang Li, Zhou Lu, Xinzhao Wang, and Rui Yang.
\newblock Adaptive online learning of quantum states.
\newblock {\em Quantum}, 8:1471, 2024.

\bibitem{lumbreras2022multi}
Josep Lumbreras, Erkka Haapasalo, and Marco Tomamichel.
\newblock Multi-armed quantum bandits: Exploration versus exploitation when
  learning properties of quantum states.
\newblock {\em Quantum}, 6:749, 2022.

\bibitem{aaronson2007learnability}
Scott Aaronson.
\newblock The learnability of quantum states.
\newblock {\em Proceedings of the Royal Society A: Mathematical, Physical and
  Engineering Sciences}, 463(2088):3089--3114, 2007.

\bibitem{mohri2018foundations}
Mehryar Mohri, Afshin Rostamizadeh, and Ameet Talwalkar.
\newblock {\em Foundations of machine learning}.
\newblock MIT press, 2018.

\bibitem{aaronson2018shadow}
Scott Aaronson.
\newblock Shadow tomography of quantum states.
\newblock In {\em Proceedings of the 50th Annual ACM SIGACT Symposium on Theory
  of Computing}, pages 325--338, 2018.

\bibitem{torlai2018neural}
Giacomo Torlai, Guglielmo Mazzola, Juan Carrasquilla, Matthias Troyer, Roger
  Melko, and Giuseppe Carleo.
\newblock Neural-network quantum state tomography.
\newblock {\em Nature Physics}, 14(5):447--450, 2018.

\bibitem{gao2018experimental}
Jun Gao, Lu-Feng Qiao, Zhi-Qiang Jiao, Yue-Chi Ma, Cheng-Qiu Hu, Ruo-Jing Ren,
  Ai-Lin Yang, Hao Tang, Man-Hong Yung, and Xian-Min Jin.
\newblock Experimental machine learning of quantum states.
\newblock {\em Physical Review Letters}, 120(24):240501, 2018.

\bibitem{zhu2022flexible}
Yan Zhu, Ya-Dong Wu, Ge~Bai, Dong-Sheng Wang, Yuexuan Wang, and Giulio
  Chiribella.
\newblock Flexible learning of quantum states with generative query neural
  networks.
\newblock {\em Nature Communications}, 13(1):6222, 2022.

\bibitem{wang2022predicting}
Haoxiang Wang, Maurice Weber, Josh Izaac, and Cedric Yen-Yu Lin.
\newblock Predicting properties of quantum systems with conditional generative
  models.
\newblock {\em arXiv preprint arXiv:2211.16943}, 2022.

\bibitem{carrasquilla2019reconstructing}
Juan Carrasquilla, Giacomo Torlai, Roger~G Melko, and Leandro Aolita.
\newblock Reconstructing quantum states with generative models.
\newblock {\em Nature Machine Intelligence}, 1(3):155--161, 2019.

\bibitem{torlai2019integrating}
Giacomo Torlai, Brian Timar, Evert~PL Van~Nieuwenburg, Harry Levine, Ahmed
  Omran, Alexander Keesling, Hannes Bernien, Markus Greiner, Vladan
  Vuleti{\'c}, Mikhail~D Lukin, et~al.
\newblock Integrating neural networks with a quantum simulator for state
  reconstruction.
\newblock {\em Physical Review Letters}, 123(23):230504, 2019.

\bibitem{xin2019local}
Tao Xin, Sirui Lu, Ningping Cao, Galit Anikeeva, Dawei Lu, Jun Li, Guilu Long,
  and Bei Zeng.
\newblock Local-measurement-based quantum state tomography via neural networks.
\newblock {\em npj Quantum Information}, 5(1):109, 2019.

\bibitem{ahmed2021quantum}
Shahnawaz Ahmed, Carlos~S{\'a}nchez Mu{\~n}oz, Franco Nori, and Anton~Frisk
  Kockum.
\newblock Quantum state tomography with conditional generative adversarial
  networks.
\newblock {\em Physical Review Letters}, 127(14):140502, 2021.

\bibitem{quek2021adaptive}
Yihui Quek, Stanislav Fort, and Hui~Khoon Ng.
\newblock Adaptive quantum state tomography with neural networks.
\newblock {\em npj Quantum Information}, 7(1):105, 2021.

\bibitem{morawetz2021u}
Stewart Morawetz, Isaac~JS De~Vlugt, Juan Carrasquilla, and Roger~G Melko.
\newblock U (1)-symmetric recurrent neural networks for quantum state
  reconstruction.
\newblock {\em Physical Review A}, 104(1):012401, 2021.

\bibitem{cha2021attention}
Peter Cha, Paul Ginsparg, Felix Wu, Juan Carrasquilla, Peter~L McMahon, and
  Eun-Ah Kim.
\newblock Attention-based quantum tomography.
\newblock {\em Machine Learning: Science and Technology}, 3(1):01LT01, 2021.

\bibitem{lange2022adaptive}
Hannah Lange, Matja{\v{z}} Kebri{\v{c}}, Maximilian Buser, Ulrich
  Schollw{\"o}ck, Fabian Grusdt, and Annabelle Bohrdt.
\newblock Adaptive quantum state tomography with active learning.
\newblock {\em Quantum}, 7:1129, 2023.

\bibitem{koutny2022neural}
Dominik Koutn{\`y}, Libor Motka, Zden{\v{e}}k Hradil, Jaroslav
  {\v{R}}eh{\'a}{\v{c}}ek, and Luis~L S{\'a}nchez-Soto.
\newblock Neural-network quantum state tomography.
\newblock {\em Physical Review A}, 106(1):012409, 2022.

\bibitem{schmale2022efficient}
Tobias Schmale, Moritz Reh, and Martin G{\"a}rttner.
\newblock Efficient quantum state tomography with convolutional neural
  networks.
\newblock {\em npj Quantum Information}, 8(1):115, 2022.

\bibitem{zhang2022efficient}
Yuan-Hang Zhang and Massimiliano Di~Ventra.
\newblock Efficient quantum state tomography with mode-assisted training.
\newblock {\em Physical Review A}, 106(4):042420, 2022.

\bibitem{an2023unified}
Zheng An, Jiahui Wu, Muchun Yang, DL~Zhou, and Bei Zeng.
\newblock Unified quantum state tomography and {Hamiltonian} learning: A
  language-translation-like approach for quantum systems.
\newblock {\em Physical Review Applied}, 21(1):014037, 2024.

\bibitem{wu2023quantum}
Ya-Dong Wu, Yan Zhu, Ge~Bai, Yuexuan Wang, and Giulio Chiribella.
\newblock Quantum similarity testing with convolutional neural networks.
\newblock {\em Physical Review Letters}, 130(21):210601, 2023.

\bibitem{zhang2021direct}
Xiaoqian Zhang, Maolin Luo, Zhaodi Wen, Qin Feng, Shengshi Pang, Weiqi Luo, and
  Xiaoqi Zhou.
\newblock Direct fidelity estimation of quantum states using machine learning.
\newblock {\em Physical Review Letters}, 127(13):130503, 2021.

\bibitem{smith2021efficient}
Alistair W.~R. Smith, Johnnie Gray, and M.~S. Kim.
\newblock Efficient quantum state sample tomography with basis-dependent neural
  networks.
\newblock {\em PRX Quantum}, 2:020348, Jun 2021.

\bibitem{zhong2022quantum}
Lu~Zhong, Chu Guo, and Xiaoting Wang.
\newblock Quantum state tomography inspired by language modeling.
\newblock {\em arXiv preprint arXiv:2212.04940}, 2022.

\bibitem{lohani2023dimension}
Sanjaya Lohani, Sangita Regmi, Joseph~M Lukens, Ryan~T Glasser, Thomas~A
  Searles, and Brian~T Kirby.
\newblock Dimension-adaptive machine learning-based quantum state
  reconstruction.
\newblock {\em Quantum Machine Intelligence}, 5(1):1, 2023.

\bibitem{nguyen2022optimizing}
H~Chau Nguyen, Jan~Lennart B{\"o}nsel, Jonathan Steinberg, and Otfried
  G{\"u}hne.
\newblock Optimizing shadow tomography with generalized measurements.
\newblock {\em Physical Review Letters}, 129(22):220502, 2022.

\bibitem{zhou2023performance}
You Zhou and Qing Liu.
\newblock Performance analysis of multi-shot shadow estimation.
\newblock {\em Quantum}, 7:1044, 2023.

\bibitem{ippoliti2023classical}
Matteo Ippoliti.
\newblock Classical shadows based on locally-entangled measurements.
\newblock {\em Quantum}, 8:1293, 2024.

\bibitem{huang2021efficient}
Hsin-Yuan Huang, Richard Kueng, and John Preskill.
\newblock Efficient estimation of {Pauli} observables by derandomization.
\newblock {\em Physical Review Letters}, 127(3):030503, 2021.

\bibitem{nakaji2023measurement}
Kouhei Nakaji, Suguru Endo, Yuichiro Matsuzaki, and Hideaki Hakoshima.
\newblock Measurement optimization of variational quantum simulation by
  classical shadow and derandomization.
\newblock {\em Quantum}, 7:995, 2023.

\bibitem{huang2022quantum}
Hsin-Yuan Huang, Michael Broughton, Jordan Cotler, Sitan Chen, Jerry Li, Masoud
  Mohseni, Hartmut Neven, Ryan Babbush, Richard Kueng, John Preskill, et~al.
\newblock Quantum advantage in learning from experiments.
\newblock {\em Science}, 376(6598):1182--1186, 2022.

\bibitem{huang2022provably}
Hsin-Yuan Huang, Richard Kueng, Giacomo Torlai, Victor~V Albert, and John
  Preskill.
\newblock Provably efficient machine learning for quantum many-body problems.
\newblock {\em Science}, 377(6613):eabk3333, 2022.

\bibitem{hu2022logical}
Hong-Ye Hu, Ryan LaRose, Yi-Zhuang You, Eleanor Rieffel, and Zhihui Wang.
\newblock Logical shadow tomography: Efficient estimation of error-mitigated
  observables.
\newblock {\em arXiv preprint arXiv:2203.07263}, 2022.

\bibitem{jnane2023quantum}
Hamza Jnane, Jonathan Steinberg, Zhenyu Cai, H~Chau Nguyen, and B{\'a}lint
  Koczor.
\newblock Quantum error mitigated classical shadows.
\newblock {\em PRX Quantum}, 5(1):010324, 2024.

\bibitem{seif2023shadow}
Alireza Seif, Ze-Pei Cian, Sisi Zhou, Senrui Chen, and Liang Jiang.
\newblock Shadow distillation: Quantum error mitigation with classical shadows
  for near-term quantum processors.
\newblock {\em PRX Quantum}, 4(1):010303, 2023.

\bibitem{sack2022avoiding}
Stefan~H Sack, Raimel~A Medina, Alexios~A Michailidis, Richard Kueng, and
  Maksym Serbyn.
\newblock Avoiding barren plateaus using classical shadows.
\newblock {\em PRX Quantum}, 3(2):020365, 2022.

\bibitem{boyd2022training}
Gregory Boyd and B{\'a}lint Koczor.
\newblock Training variational quantum circuits with {CoVaR}: Covariance root
  finding with classical shadows.
\newblock {\em Physical Review X}, 12(4):041022, 2022.

\bibitem{du2022demystify}
Yuxuan Du, Yibo Yang, Dacheng Tao, and Min-Hsiu Hsieh.
\newblock Problem-dependent power of quantum neural networks on multiclass
  classification.
\newblock {\em Physical Review Letters}, 131(14):140601, 2023.

\bibitem{ma2023attention}
H~Ma, Z~Sun, D~Dong, C~Chen, and H~Rabitz.
\newblock Tomography of quantum states from structured measurements via
  quantum-aware transformer.
\newblock {\em IEEE Transactions on Cybernetics}, 55(6):2571--2582, 2025.

\bibitem{nir2020machine}
Amit Nir, Eran Sela, Roy Beck, and Yohai Bar-Sinai.
\newblock Machine-learning iterative calculation of entropy for physical
  systems.
\newblock {\em Proceedings of the National Academy of Sciences},
  117(48):30234--30240, 2020.

\bibitem{xiao2022intelligent}
Tailong Xiao, Jingzheng Huang, Hongjing Li, Jianping Fan, and Guihua Zeng.
\newblock Intelligent certification for quantum simulators via machine
  learning.
\newblock {\em npj Quantum Information}, 8(1):138, 2022.

\bibitem{qian2024multimodal}
Yang Qian, Yuxuan Du, Zhenliang He, Min-Hsiu Hsieh, and Dacheng Tao.
\newblock Multimodal deep representation learning for quantum cross-platform
  verification.
\newblock {\em Physical Review Letters}, 133(13):130601, 2024.

\bibitem{kliesch2021theory}
Martin Kliesch and Ingo Roth.
\newblock Theory of quantum system certification.
\newblock {\em PRX Quantum}, 2(1):010201, 2021.

\bibitem{nielsen2010quantum}
Michael~A Nielsen and Isaac~L Chuang.
\newblock {\em Quantum computation and quantum information}.
\newblock Cambridge University Press, 2010.

\bibitem{elben2020cross}
Andreas Elben, Beno{\^\i}t Vermersch, Rick Van~Bijnen, Christian Kokail, Tiff
  Brydges, Christine Maier, Manoj~K Joshi, Rainer Blatt, Christian~F Roos, and
  Peter Zoller.
\newblock Cross-platform verification of intermediate scale quantum devices.
\newblock {\em Physical Review Letters}, 124(1):010504, 2020.

\bibitem{huang2022learning}
Hsin-Yuan Huang.
\newblock Learning quantum states from their classical shadows.
\newblock {\em Nature Reviews Physics}, 4(2):81--81, 2022.

\bibitem{vaswani2017attention}
Ashish Vaswani, Noam Shazeer, Niki Parmar, Jakob Uszkoreit, Llion Jones,
  Aidan~N Gomez, {\L}ukasz Kaiser, and Illia Polosukhin.
\newblock Attention is all you need.
\newblock {\em Advances in Neural Information Processing Systems}, 30, 2017.

\bibitem{dosovitskiy2020image}
Alexey Dosovitskiy, Lucas Beyer, Alexander Kolesnikov, Dirk Weissenborn,
  Xiaohua Zhai, Thomas Unterthiner, Mostafa Dehghani, Matthias Minderer, Georg
  Heigold, Sylvain Gelly, Jakob Uszkoreit, and Neil Houlsby.
\newblock An image is worth 16x16 words: Transformers for image recognition at
  scale.
\newblock In {\em International Conference on Learning Representations}, 2021.

\bibitem{Devlin2019BERT}
Jacob Devlin, Ming{-}Wei Chang, Kenton Lee, and Kristina Toutanova.
\newblock {BERT}: Pre-training of deep bidirectional transformers for language
  understanding.
\newblock In Jill Burstein, Christy Doran, and Thamar Solorio, editors, {\em
  Proceedings of the 2019 Conference of the North American Chapter of the
  Association for Computational Linguistics: Human Language Technologies,
  {NAACL-HLT} 2019, Minneapolis, MN, USA, June 2-7, 2019, Volume 1 (Long and
  Short Papers)}, pages 4171--4186. Association for Computational Linguistics,
  2019.

\bibitem{banaszek1999maximum}
Konrad Banaszek, GM~D’ariano, MGA Paris, and MF~Sacchi.
\newblock Maximum-likelihood estimation of the density matrix.
\newblock {\em Physical Review A}, 61(1):010304, 1999.

\bibitem{lu2018separability}
Sirui Lu, Shilin Huang, Keren Li, Jun Li, Jianxin Chen, Dawei Lu, Zhengfeng Ji,
  Yi~Shen, Duanlu Zhou, and Bei Zeng.
\newblock Separability-entanglement classifier via machine learning.
\newblock {\em Physical Review A}, 98(1), 2018.

\bibitem{zeiler2010deconvolutional}
Matthew~D Zeiler, Dilip Krishnan, Graham~W Taylor, and Rob Fergus.
\newblock Deconvolutional networks.
\newblock In {\em 2010 IEEE Computer Society Conference on Computer Vision and
  Pattern Recognition}, pages 2528--2535. IEEE, 2010.

\bibitem{ulyanov2016instance}
Dmitry Ulyanov, Andrea Vedaldi, and Victor Lempitsky.
\newblock Instance normalization: The missing ingredient for fast stylization.
\newblock {\em arXiv preprint arXiv:1607.08022}, 2016.

\bibitem{xu2019understanding}
Jingjing Xu, Xu~Sun, Zhiyuan Zhang, Guangxiang Zhao, and Junyang Lin.
\newblock Understanding and improving layer normalization.
\newblock {\em Advances in Neural Information Processing Systems}, 32, 2019.

\bibitem{monz201114}
Thomas Monz, Philipp Schindler, Julio~T Barreiro, Michael Chwalla, Daniel Nigg,
  William~A Coish, Maximilian Harlander, Wolfgang H{\"a}nsel, Markus Hennrich,
  and Rainer Blatt.
\newblock 14-qubit entanglement: Creation and coherence.
\newblock {\em Physical Review Letters}, 106(13):130506, 2011.

\bibitem{he2015spatial}
Kaiming He, Xiangyu Zhang, Shaoqing Ren, and Jian Sun.
\newblock Spatial pyramid pooling in deep convolutional networks for visual
  recognition.
\newblock {\em IEEE Transactions on Pattern Analysis and Machine Intelligence},
  37(9):1904--1916, 2015.

\bibitem{koutny2023deep}
Dominik Koutn{\`y}, Laia Gin{\'e}s, Magdalena Mocza{\l}a-Dusanowska, Sven
  H{\"o}fling, Christian Schneider, Ana Predojevi{\'c}, and Miroslav
  Je{\v{z}}ek.
\newblock Deep learning of quantum entanglement from incomplete measurements.
\newblock {\em Science Advances}, 9(29):eadd7131, 2023.

\bibitem{carleo2019machine}
Giuseppe Carleo, Ignacio Cirac, Kyle Cranmer, Laurent Daudet, Maria Schuld,
  Naftali Tishby, Leslie Vogt-Maranto, and Lenka Zdeborov{\'a}.
\newblock Machine learning and the physical sciences.
\newblock {\em Reviews of Modern Physics}, 91(4):045002, 2019.

\bibitem{bergholm2018pennylane}
Ville Bergholm, Josh Izaac, Maria Schuld, Christian Gogolin, Shahnawaz Ahmed,
  Vishnu Ajith, M~Sohaib Alam, Guillermo Alonso-Linaje, B~AkashNarayanan, Ali
  Asadi, et~al.
\newblock {PennyLane}: Automatic differentiation of hybrid quantum-classical
  computations.
\newblock {\em arXiv preprint arXiv:1811.04968}, 2018.

\bibitem{pastaq}
Giacomo Torlai and Matthew Fishman.
\newblock \mbox{PastaQ}: A package for simulation, tomography and analysis of
  quantum computers, 2020.

\bibitem{van2008visualizing}
Laurens Van~der Maaten and Geoffrey Hinton.
\newblock Visualizing data using t-{SNE}.
\newblock {\em Journal of Machine Learning Research}, 9(11), 2008.

\bibitem{bertoni2022shallow}
Christian Bertoni, Jonas Haferkamp, Marcel Hinsche, Marios Ioannou, Jens
  Eisert, and Hakop Pashayan.
\newblock Shallow shadows: Expectation estimation using low-depth random
  {Clifford} circuits.
\newblock {\em Physical Review Letters}, 133(2):020602, 2024.

\bibitem{hadfield2022measurements}
Charles Hadfield, Sergey Bravyi, Rudy Raymond, and Antonio Mezzacapo.
\newblock Measurements of quantum {Hamiltonians} with locally-biased classical
  shadows.
\newblock {\em Communications in Mathematical Physics}, 391(3):951--967, 2022.

\bibitem{struchalin2021experimental}
GI~Struchalin, Ya~A Zagorovskii, EV~Kovlakov, SS~Straupe, and SP~Kulik.
\newblock Experimental estimation of quantum state properties from classical
  shadows.
\newblock {\em PRX Quantum}, 2(1):010307, 2021.

\bibitem{wei2023neural}
Victor Wei, WA~Coish, Pooya Ronagh, and Christine~A Muschik.
\newblock Neural-shadow quantum state tomography.
\newblock {\em Physical Review Research}, 6(2):023250, 2024.

\bibitem{becker2022classical}
Simon Becker, Nilanjana Datta, Ludovico Lami, and Cambyse Rouz{\'e}.
\newblock Classical shadow tomography for continuous variables quantum systems.
\newblock {\em IEEE Transactions on Information Theory}, 70(5):3427--3452,
  2024.

\bibitem{preskill2018quantum}
John Preskill.
\newblock Quantum computing in the {NISQ} era and beyond.
\newblock {\em Quantum}, 2:79, 2018.

\bibitem{cerezo2021variational}
Marco Cerezo, Andrew Arrasmith, Ryan Babbush, Simon~C Benjamin, Suguru Endo,
  Keisuke Fujii, Jarrod~R McClean, Kosuke Mitarai, Xiao Yuan, Lukasz Cincio,
  et~al.
\newblock Variational quantum algorithms.
\newblock {\em Nature Reviews Physics}, 3(9):625--644, 2021.

\bibitem{bharti2021noisy}
Kishor Bharti, Alba Cervera-Lierta, Thi~Ha Kyaw, Tobias Haug, Sumner
  Alperin-Lea, Abhinav Anand, Matthias Degroote, Hermanni Heimonen, Jakob~S
  Kottmann, Tim Menke, et~al.
\newblock Noisy intermediate-scale quantum algorithms.
\newblock {\em Reviews of Modern Physics}, 94(1):015004, 2022.

\bibitem{tian2023recent}
Jinkai Tian, Xiaoyu Sun, Yuxuan Du, Shanshan Zhao, Qing Liu, Kaining Zhang, Wei
  Yi, Wanrong Huang, Chaoyue Wang, Xingyao Wu, et~al.
\newblock Recent advances for quantum neural networks in generative learning.
\newblock {\em IEEE Transactions on Pattern Analysis and Machine Intelligence},
  2023.

\bibitem{li2022recent}
Weikang Li and Dong-Ling Deng.
\newblock Recent advances for quantum classifiers.
\newblock {\em Science China Physics, Mechanics \& Astronomy}, 65(2):1--23,
  2022.

\bibitem{bennewitz2022neural}
Elizabeth~R Bennewitz, Florian Hopfmueller, Bohdan Kulchytskyy, Juan
  Carrasquilla, and Pooya Ronagh.
\newblock Neural error mitigation of near-term quantum simulations.
\newblock {\em Nature Machine Intelligence}, 4(7):618--624, 2022.

\end{thebibliography}
%

 \newpage
 
\appendices

\onecolumn
\begin{center}
\large{\textbf{Supplemental Material for ``ShadowNet for Data-Centric Quantum System Learning''}}	
\end{center}

\renewcommand{\thesection}{\Alph{section}}
\numberwithin{equation}{section}
\renewcommand\thefigure{\thesection.\arabic{figure}}    

\appendices
 
\section{Faithfulness of ShadowNet}\label{append:sec:faith-evaluation}
In this section, we investigate how the number of shots used in classical shadows affects the predictive faith of ShadowNet, mainly focusing on Pauli-based measurements to align with the main text. We end this section by explaining how to enhance the predictive faith when Pauli-based measurements incur a large estimation error bound.

Following the mechanism of ShadowNet, quantifying faithfulness involves determining the estimation error of classical shadows. In this regard, we can leverage the following two lemmas developed by \cite{huang2020predicting} to complete the analysis.
\begin{lemma}[Theorem 1, adapted from \cite{huang2020predicting}]\label{lemma:shadowres-1}
	Fix a measurement primitive $\mathcal{U}$, a collection $O_1,...,O_L$ of $2^N\times 2^N$ Hermitian matrices and accuracy parameters $\epsilon, \delta\in (0, 1)$. Set
	\begin{equation}
		K = 2\log(2L/\delta)~\text{and}~R=\frac{34}{\epsilon^2}\max_{1\leq i \leq L}\left\|O_i-\frac{\Tr(O_i)}{2^N}\mathbb{I}\right\|^2_{\text{shadow}}, 
	\end{equation}
	where $\|\cdot\|_{\text{shadow}}$ refers to the shadow norm. Then, a collection of $M=KR$ independent classical shadow allows for accurately predicting all features via median of means prediction
	\begin{equation}
		|\hat{o}_i(R,K)-\Tr(O_i\rho)|\leq \epsilon,~\forall i \in [L], 
	\end{equation}
	with probability $1-\delta$. 
\end{lemma}
\begin{lemma}[Lemma 3, adapted from  \cite{huang2020predicting}]\label{lemma:shadowres-2}
	Let $O$ be a single $k$-local Pauli observable, e.g., $O=P_{p_1}\otimes\cdots P_{p_k}\otimes\mathbb{I}_{2^{N-k}}$, where $P_j\in \{X,Y,Z\}$. Then, $\|O\|_{\text{shadow}}^2=3^k$, for any choice of the k qubits where nontrivial Pauli matrices act. For observables written as linear combinations of Pauli strings, we apply this result term by term.
\end{lemma}
The first lemma quantifies the tradeoff between the approximation error of classical shadows $\epsilon$ and the number of measurements $M$. The second lemma quantifies the upper bound of the shadow norm when the Pauli-based measurements are adopted.

\medskip 
\noindent\textbf{Ground energy estimation}. Suppose the specified $N$-qubit spin system is $k$-local, i.e., the corresponding Hamiltonian is $H=\sum_{i=1}^L \alpha_i P_{p_1}^{(i)}\otimes P_{p_2}^{(i)} \cdots \otimes P_{p_k}^{(i)}\otimes \mathbb{I}_{2^{N-k}}$ with $P_j\in \{X,Y,Z\}$. For simplicity, each Pauli term  is denoted by $\vec{P}^{(i)}=P_{p_1}^{(i)}\otimes P_{p_2}^{(i)} \cdots \otimes P_{p_k}^{(i)}\otimes \mathbb{I}_{2^{N-k}}$ for $\forall i\in[L]$. Let $\rho$ be the ground state of $H$ and $\hatrho$ be the corresponding shadows under $M$ measurements. According to the explicit form of $H$, the estimation error is upper bounded by
\begin{align}
\label{append:eqn:coro-shadow-Hamiltonian-1}
	|\Tr(\hatrho H) - \Tr(\rho H)| 
	& = \left|\sum_{i=1}^L \alpha_i \Big(\Tr(\hatrho \vec{P}^{(i)})-\Tr(\rho P^{(i)})\Big) \right| \nonumber\\
 & \leq \sum_{i=1}^L |\alpha_i||\Tr(\hatrho \vec{P}^{(i)})-\Tr(\rho P^{(i)})|. 
\end{align}
Define $\epsilon_i:=|\Tr(\hatrho \vec{P}^{(i)})-\Tr(\rho P^{(i)})|$ as the estimation error of the $i$-th term for $\forall i \in [L]$. Lemma \ref{lemma:shadowres-1} indicates that when $K=2\log(2L/\delta)$, with probability $1-\delta$,
\begin{equation}
	\epsilon_i^2  \leq \frac{34}{R}\max_{1\leq i \leq L} \left\|\vec{P}^{(i)}-\frac{\Tr(\vec{P}^{(i)})}{2^N}\mathbb{I} \right\|^2_{\text{shadow}},~ \forall i\in[L]. 
\end{equation}
Then Lemma \ref{lemma:shadowres-2} gives 
\begin{equation}
	\max_{1\leq i \leq L} \left\|\vec{P}^{(i)}-\frac{\Tr(\vec{P}^{(i)})}{2^N}\mathbb{I} \right\|^2_{\text{shadow}} = \max_{1\leq i \leq L} \left\|\vec{P}^{(i)} \right\|^2_{\text{shadow}} \leq 3^k,
\end{equation}
where the first equality uses the traceless property of Pauli operators. 

Combining the above two equations, we obtain
\begin{equation}\label{append:eqn:coro-shadow-Hamiltonian-2}
	\epsilon_i^2 \leq \frac{34\cdot 3^k}{R}.
\end{equation} 
Supporting by this result and Eq.~(\ref{append:eqn:coro-shadow-Hamiltonian-1}), we can conclude that with probability $1-\delta$, the estimation error of classical shadows in learning spin systems is supper bounded by 
\begin{equation}\label{eqn:append:faith-ground-energy}
	|\Tr(\hatrho H) - \Tr(\rho H)| \leq \sum_{i=1}^L |\alpha_i| \sqrt{\frac{34\cdot 3^k}{R}}.
\end{equation}

For illustration, we employ the result in Eq.~(\ref{eqn:append:faith-ground-energy}) to analyze the predictive faith of ShadowNet in numerical simulations. In the task of reconstructing ground states of TFIM, we have $N=5$, $L=9$, and $k=1, 2$. Let $K=\left\lceil 2\log\frac{2L}{0.166}\right\rceil=10$. When $M=10000$, with probability at least $83.4\%$, we choose  $R=M/K=1000$, the estimation error is upper bounded by $\epsilon \le 4\cdot 0.5\cdot \sqrt{\frac{34\cdot 9}{1000}} + 5\cdot \sqrt{\frac{34\cdot 3}{1000}} \approx 2.7$. When $M=10$, we have $R=M/K=1$. Therefore, with probability at least $83.4\%$, the estimation error is upper bounded by $\epsilon \leq  4\cdot 0.5\cdot \sqrt{34\cdot 9} + 5\cdot \sqrt{34\cdot 3} \approx 85.48.$

In the task of jointly reconstructing the ground states of TFIM and XXZ models, we further quantify the estimation error of XXZ model and preserve the maximum one. Concretely, we have  $N=5$, $L=12$, and $k=2$ for all local terms. Let $K=\left\lceil 2\log\frac{2L}{0.166}\right\rceil=10$. When $M=10000$, with probability at least $83.4\%$, we choose $R=M/K=1000$, and the estimation error for the ground energy of the XXZ model is upper bounded by $\epsilon \le 12\cdot 3\cdot \sqrt{\frac{34\cdot 9}{1000}} \approx 19.91$. When $M=10$, we have $R=M/K=1$. Therefore, with probability at least $83.4\%$, the estimation error for the ground energy of the XXZ model is upper bounded by $\epsilon \le 12\cdot 3\cdot \sqrt{34\cdot 9} \approx 629.75$. Since this bound is larger than the corresponding TFIM bound, in the joint learning task of TFIM and XXZ models, the error bound of classical shadows is $629.75$ ($19.91$) when $M=10$ ($M=10000$).

\begin{figure*}[h]
\centering
\includegraphics[width=0.98\textwidth]{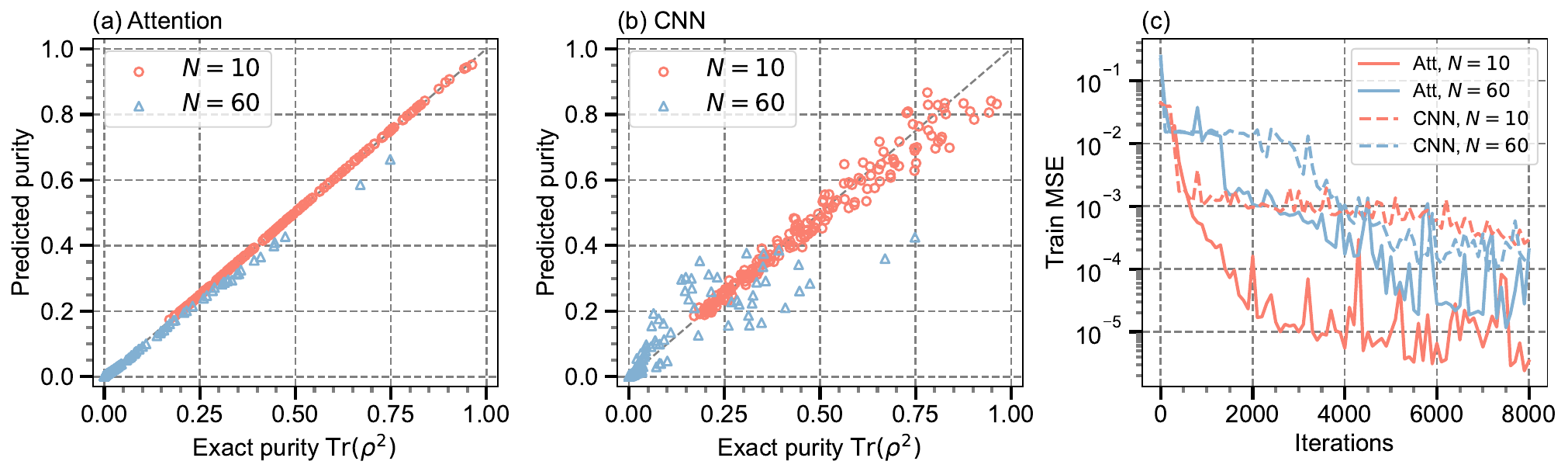}
\caption{\small{\textbf{Predicting the nonlinear property $\Tr(\rho^2)$ of noisy GHZ states with ShadowNet.} (a) The predicted versus exact purity on $200$ unseen test states for the attention-based implementation with $N=10$ (red circles) and $N=60$ (blue triangles) qubits, where the dashed line denotes the perfect prediction. (b) The same for the convolutional implementation. (c) The training dynamics, measured by the MSE on the training dataset, of the four settings. The scatter plots in (a) and (b) show one representative seed, and the quantitative results averaged over $3$ seeds are reported in Table~\ref{tab:purity-rebuttal}.}}
\label{fig:purity-rebuttal}
\end{figure*}

\medskip
\noindent\textbf{Direct fidelity estimation}. The above analysis can be effectively generalized to the task of direct fidelity estimation (DFE). In this task, the Hamiltonian $H$ in Eq.~(\ref{append:eqn:coro-shadow-Hamiltonian-1}) refers to the ideal-state projector $\ket{\psi}\bra{\psi}$, and $\rho$ corresponds to its noisy version generated by near-term quantum computers. Denote the decomposition of $\ket{\psi}$ under Pauli-basis as
\begin{equation}\label{append:eqn:DFE-pauli-decompose}
	\ket{\psi}\bra{\psi} = \sum_{i=1}^{4^N} \bm{\gamma}_i \vec{P}^{(i)},  
\end{equation}  
where $\vec{P}^{(i)}\in \{X,Y,Z,\mathbb{I}_2\}^{\otimes N}$ refers to the $i$-th Pauli term and $\bm{\gamma}_i= 2^{-N}\Tr(\vec{P}^{(i)}\ket{\psi}\bra{\psi})$. The sparsity of the coefficients $\bm{\gamma}=[\bm{\gamma}_1, ..., \bm{\gamma}_i, ..., \bm{\gamma}_{4^N}]$ amounts to $\|\bm{\gamma}\|_0$. 

Denote $L=\|\bm{\gamma}\|_0$. Suppose that the locality of $\vec{P}^{(i)}$ in Eq.~(\ref{append:eqn:DFE-pauli-decompose}) is at most $k$ with $\bm{\gamma}_i\neq 0$. As with the analysis of spin systems, we can conclude that with probability $1-\delta$, the estimation error of classical shadows in DFE is upper bounded by 
\begin{equation}\label{append:eqn:shadow-faith-dfe-bound}
	|\Tr(\hatrho \ket{\psi}\bra{\psi}) - \Tr(\rho \ket{\psi}\bra{\psi})| \leq \sum_{i=1}^L |\bm{\gamma}_i| \sqrt{\frac{34\cdot 3^k}{R}}.
\end{equation} 

For illustration, we next analyze the predictive faith of ShadowNet in numerical simulations. For an $N$-qubit GHZ state with $\ket{\GHZ}=(\ket{0}^{\otimes N} + \ket{1}^{\otimes N})/\sqrt{2}$, its Pauli-decomposition yields 
\begin{eqnarray}
	  \ket{\GHZ}\bra{\GHZ}  = && \frac{1}{2^{N}}\sum_{t=0}^{\lfloor N/2\rfloor} \Big[\sum_{\bm{\pi}}S_{\bm{\pi}}(\mathbb{I}^{\otimes (N-2t)}\otimes Z^{\otimes 2t})  \nonumber\\
&& + (-1)^t\sum_{\bm{\pi}}S_{\bm{\pi}}(X^{\otimes (N-2t)}\otimes Y^{\otimes 2t})\Big], \nonumber
\end{eqnarray}  
where $S_{\bm{\pi}}$ denotes the permutation operator for some permutation $\bm{\pi}:\{1,...,N\}\rightarrow \{1, ..., N\}$. When $N=60$, the total number of Pauli basis is $L = 2 \times (1 + C_{60}^2 + C_{60}^4 + ... + C_{60}^{60})=2 \times 2^{59}=2^{60}$ with $|\bm{\gamma}_i|=1/2^{60}$ for $\forall i \in [L]$. According to Eq.~(\ref{append:eqn:shadow-faith-dfe-bound}), the estimation error is upper bounded by $\epsilon=\frac{1}{2^{60}}\sum_{i=1}^{2^{60}}\sqrt{34\cdot 3^k /R}\leq \sqrt{34 \cdot 3^{60} / R}$. This error bound hints that classical shadows with  Pauli-based measurements are not a proper choice to evaluate the faithfulness of ShadowNet, due to the large value of the locality $k$.

\medskip 
\noindent\textbf{Enhancement of predictive faith}. ShadowNet adopts random Pauli-based measurements to collect classical shadows, attributed to its implementable property and the capability of providing the form of local shadows. However, for some important observables, the Pauli expansion can contain exponentially many nonzero terms, and the locality of these Pauli terms can scale linearly with the system size. In such cases, Pauli-based classical shadows can require a prohibitively large number of measurements. To alleviate this issue, a possible solution is adopting other random measurement protocols such as random Clifford measurements rather than Pauli-based measurements \textit{at the inference stage} to dramatically suppress the estimation error bound. The feasibility of using Clifford measurements comes from two aspects. First, ShadowNet does not require the measurement group to be the same in inference and the evaluation of predictive faith. Moreover, although Clifford measurement is difficult to implement on modern quantum machines, theoretical results have shown that it requests very few measurements to accurately estimate $\Tr(\rho \ket{\psi}\bra{\psi})$ for a specific class of $\ket{\psi}$. For instance,  Clifford measurements can be exploited in the inference stage to dramatically reduce the estimation error in the numerical task related to DFE. As proved in \cite{huang2020predicting}, when $M=2000$ ($M=20000$), with probability at least $0.95$, the estimation error is upper bounded by  $\epsilon\le 0.64$ ($\epsilon\le 0.20$).

\section{Experiments for nonlinear-property prediction}

The experiments in the main text focus on linear-property prediction. For completeness, we further apply ShadowNet to purity prediction, a representative nonlinear-property prediction task. Below, we first describe the experimental setup, followed by the hyperparameter settings and an analysis of the results.

\medskip
\noindent\textbf{Experiment setup}. The \textit{purity} $\Tr(\rho^2)$ of noisy quantum states $\rho$ is a prototypical nonlinear property $\Tr(\rho^k O)$, with setting $k=2$ and $O=\mathbb{I}$. Here the state $\rho$ to be explored is collected from the noisy GHZ dataset, which is the same as that used in the \textit{direct fidelity estimation} (DFE) task in the main text (see Sec.~4.2). In particular, each training example corresponds to a noisy $N$-qubit GHZ state with 
\[ \rho(\bm{p})=\mathcal{E}_{\bm{p}}(\ket{\text{GHZ}}\bra{\text{GHZ}}),\] where $\ket{\text{GHZ}}=(\ket{0}^{\otimes N} + \ket{1}^{\otimes N})/\sqrt{2}$ is the $N$-qubit GHZ state, and $\bm{p}=\{p_1, p_2\}$ refers to the local depolarizing noise rate with $p_1\in[10^{-4},10^{-2}]$ for all single-qubit gates and $p_2\in[10^{-4},10^{-1}]$ for all two-qubit gates.  

We follow the procedure introduced in Sec.~3.2 in the main text to construct the training dataset $\mathcal{D}_{\text{Tr}}$. Specifically, the input feature $\bxi$ for the $i$-th training example is identical to the DFE experiment, i.e., the locally-averaged inverse snapshots constructed from $M$ randomized Pauli measurements together with $(p_1,p_2)$. Here the depolarizing noise rates $\bm{p}$ are uniformly sampled per training example. The \textit{only} change compared with the DFE task is the label, i.e., for each example we compute the exact purity $\Tr(\rho(\bm{p})^2)$ as the label instead of the fidelity. The test dataset is constructed in the same manner.

\medskip
\noindent \textbf{Hyperparameter settings}. We evaluate both implementations of ShadowNet described in Sec.~3.4 of the main text, namely the attention-based model and the convolutional model. Both models use the same hyperparameter settings as those used in the DFE tasks. Specifically, we consider $N\in\{10,60\}$, vary the number of training examples over $n\in\{20,200,800\}$, use $200$ test examples, and set the measurement budget for each state to $M=2000$ snapshots, as in the DFE task. The batch size is set to $8$, and the total number of training iterations is set to $8000$. Both models are trained by minimizing the mean squared error between the predicted and exact purities using the AdamW optimizer, with $\beta_1=0.9$, $\beta_2=0.999$, and no weight decay. The learning rate is linearly warmed up to $2\times 10^{-4}$ during the first $100$ iterations and then gradually decayed. Each setting is repeated $3$ times with independent random seeds, and we report the mean and standard deviation over these repetitions.

\begin{table}[h!]
\centering
\caption{\small{Test performance of ShadowNet on the purity prediction task.}}
\label{tab:purity-rebuttal}
\resizebox{0.98\linewidth}{!}{
\begin{tabular}{lcccc}
\toprule
 & \multicolumn{2}{c}{Attention} & \multicolumn{2}{c}{CNN} \\
 & $N=10$ & $N=60$ & $N=10$ & $N=60$ \\
\midrule
Test MSE & $(9.0\pm 5.2)\times 10^{-6}$ & $(9.1\pm 7.9)\times 10^{-5}$ & $(1.4\pm 0.1)\times 10^{-3}$ & $(3.4\pm 0.4)\times 10^{-3}$ \\
\bottomrule
\end{tabular}
}
\end{table}

\medskip
\noindent\textbf{Results and analysis}. For comprehensive, we conduct two experiments to evaluate the performance of ShadowNet in predicting the purity of noisy GHZ states.  
 
\medskip
\noindent\textit{The training dynamics between attention-based and convolutional ShadowNets}. The first experiment focuses on comparing the training dynamics of the two ShadowNet implementations by setting $n=800$.

\begin{figure}[h!]
\centering
\includegraphics[width=0.4\textwidth]{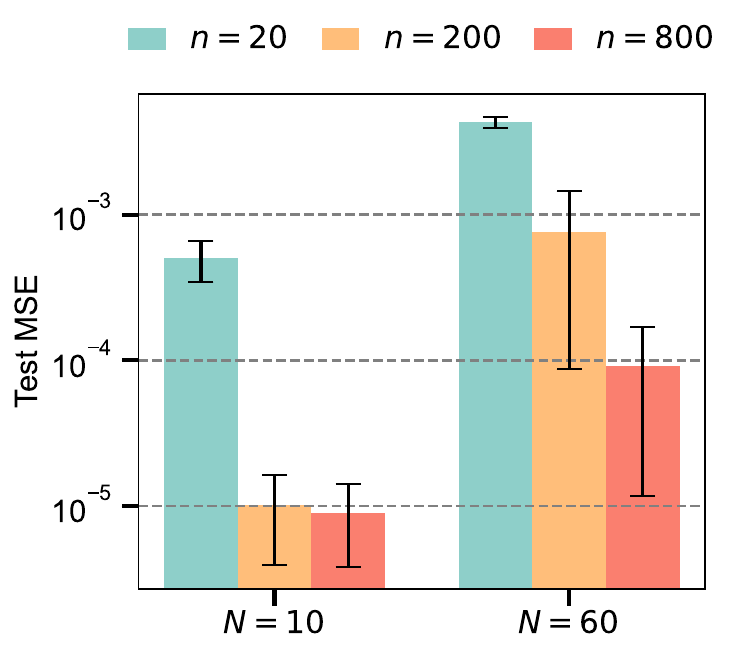}
\caption{\small{\textbf{Performance of the attention-based ShadowNet on the purity prediction task with the varied number of training examples.} The test MSE for $n\in\{20, 200, 800\}$ training examples and $N\in\{10, 60\}$ qubits, evaluated on the same $200$ test examples. The bar height and the error bar refer to the mean and the standard deviation over $3$ independent random seeds.}}
\label{fig:purity-trainsize-rebuttal}
\end{figure}

The results are summarized in Figure~\ref{fig:purity-rebuttal} and Table~\ref{tab:purity-rebuttal}. The attention-based ShadowNet achieves test mean squared errors (MSE) of $(9.0\pm 5.2)\times 10^{-6}$ for $N=10$ and $(9.1\pm 7.9)\times 10^{-5}$ for $N=60$, averaged over $3$ random seeds. The convolutional implementation attains test MSE of $(1.4\pm 0.1)\times 10^{-3}$ and $(3.4\pm 0.4)\times 10^{-3}$, respectively. 

These results show that both implementations can learn the nonlinear purity from the same dataset $\mathcal{D}_{\text{Tr}}$, while the attention-based model achieves higher accuracy. This is consistent with the linear-property prediction results reported in the main text and further supports the data-centric perspective of ShadowNet. That is, the ability to predict a new property is primarily enabled by the construction of the training dataset, whereas the choice of neural architecture mainly affects the attainable accuracy.

\medskip
\noindent\textit{The scaling behavior}. We next explore the scaling behavior of ShadowNet as the number of training examples $n$ increases. Since the attention-based ShadowNet slightly outperforms the convolutional ShadowNet, we use the former for this analysis. Specifically, we vary the number of training examples as $n\in\{20,200,800\}$ while keeping all other settings unchanged. The same $200$ test examples are used across all settings to ensure a fair comparison.

The achieved results are demonstrated in Figure~\ref{fig:purity-trainsize-rebuttal}. An observation is that a modest dataset size is sufficient to achieve strong performance. In particular, for $N=10$, increasing $n$ from $20$ to $200$ reduces the test MSE by more than one order of magnitude, which decreases from $(5.0\pm 1.6)\times 10^{-4}$ to $(1.0\pm 0.6)\times 10^{-5}$. For $N=60$, the same increase in training size reduces the error by about a factor of five, i.e., the error decreases from $(4.3\pm 0.4)\times 10^{-3}$ to $(7.7\pm 6.8)\times 10^{-4}$. Further increasing $n$ to $800$ mainly benefits the larger system. For $N=60$, the error further decreases to $(9.1\pm 7.9)\times 10^{-5}$. For $N=10$, the performance has nearly saturated, with a test mean squared error of $(9.0\pm 5.2)\times 10^{-6}$.
 
This scaling behavior echoes the dataset-size analysis of the DFE task in the main text. That is, ShadowNet achieves satisfactory performance for nonlinear property prediction once the dataset size exceeds a modest threshold.

\end{document}